\documentclass[pdflatex,sn-mathphys-sn-chicago]{sn-jnl}

\usepackage{natbib}

\usepackage{graphicx}%
\usepackage{multirow}%
\usepackage{amsmath,amssymb,amsfonts}%
\usepackage{amsthm}%
\usepackage{mathrsfs}%
\usepackage[title]{appendix}%
\usepackage{xcolor}%
\usepackage{textcomp}%
\usepackage{manyfoot}%
\usepackage{booktabs}%
\usepackage{algorithm}%
\usepackage{algorithmicx}%
\usepackage{algpseudocode}%
\usepackage{listings}%
\usepackage{subcaption}%
\usepackage{booktabs}
\usepackage{graphicx}
\usepackage{float}
\usepackage{bm}
\usepackage{enumitem}

\theoremstyle{thmstyleone}%
\theoremstyle{thmstyletwo}%

\theoremstyle{thmstylethree}%

\begin{document}

\title[Article Title]{Robust model-based clustering via mixtures of multivariate pseudo-Voigt distributions}


\author[1]{\fnm{Babak} \sur{F.\ Dehkordi}}\email{Babak.dehkordi@ubc.ca}

\author[2]{\fnm{Jeffrey} \sur{L.\ Andrews}}\email{jeff.andrews@ubc.ca}

\author[3]{\fnm{Andrew} \sur{Jirasek}}\email{andrew.jirasek@ubc.ca}

\affil[1]{\orgdiv{Department of Statistics}, \orgname{University of British Columbia-Okanagan}, \orgaddress{\city{Kelowna}, \postcode{V1V 1V7}, \state{BC}, \country{Canada}}}

\affil[2]{\orgdiv{Department of Statistics}, \orgname{University of British Columbia-Okanagan}, \orgaddress{\city{Kelowna}, \postcode{V1V 1V7}, \state{BC}, \country{Canada}}}

\affil[3]{\orgdiv{Department of Physics}, \orgname{University of British Columbia-Okanagan}, \orgaddress{\city{Kelowna}, \postcode{V1V 1V7}, \state{BC}, \country{Canada}}}


\abstract{
We propose a multivariate extension of the pseudo-Voigt profile---a weighted convex combination of Gaussian and Cauchy distributions---within a finite mixture modeling framework for robust model-based clustering and outlier detection. To ensure parsimony and coherence within clusters, shared location and scale parameters are imposed between the Gaussian and Cauchy components. 
Parameter estimation is carried out via an Expectation Maximization algorithm, with latent variables facilitating efficient likelihood-based inference. The performance of the proposed model is evaluated through simulation studies and applications to real-world data. Comparisons with established robust models, including mixtures of contaminated normal distributions, are provided to illustrate the model's clustering accuracy and outlier detection capabilities. The framework is shown to be particularly effective for data characterized by heavy-tailed behavior.}




\maketitle

\section{Introduction}\label{sec1}

Finite mixture models offer a flexible probabilistic framework for modeling complex, unexplained heterogeneity in multivariate data \citep{McLachlanPeel2000}. These models assume that observed data originate from a weighted combination of a finite number of latent component distributions, with each component corresponding to a distinct subpopulation within the overall population. Although the theoretical foundations of finite mixture models are well established, their practical implementation has been advanced by two key developments. First, improvements in computational resources have enabled the application of iterative likelihood-based estimation methods, such as the Expectation–Maximization (EM) algorithm \citep{dempster1977em}. Second, the development of parsimonious families of Gaussian mixture models (GMMs), particularly by \cite{banfield1993model}, has addressed the challenge of balancing model flexibility and parsimony to prevent over-parameterization. The use of eigen-decomposition for component covariance matrices permits a broad range of cluster shapes, volumes, and orientations while maintaining a manageable number of parameters. This approach is now foundational in contemporary model-based clustering \citep{celeux1995gaussian, fraley1998many, mcnicholas2016mixture}.

GMMs have become the standard tool for clustering and density estimation \citep{titterington1985statistical, scott1992multivariate}. Their mathematical tractability and the interpretability of the normal distribution have established them as the preferred choice across diverse scientific disciplines, including bioinformatics and image processing \citep{bouveyron2014model, mcnicholas2016mixture}. However, the use of Gaussian components assumes that underlying clusters are relatively symmetric and that the data are not affected by extreme observations.

The limitations of standard GMMs are evident when data are generated from heavy-tailed distributions or contain outliers. For heavy-tailed data, Gaussian components are often insufficient because they cannot adequately represent increased tail density, leading to over-inflated covariance estimates to accommodate extreme observations and/or additional components. To address this issue, mixtures of multivariate $t$-distributions have been proposed \citep{PeelMcLachlan2000, andrews2012model, andrews2018teigen}. The inclusion of a degrees-of-freedom parameter allows these models to accommodate heavier tails, offering a more robust framework for clusters with greater dispersion than permitted by the normal distribution.

Beyond the challenge of modeling heavy tails, a major issue in clustering is identifying atypical observations within the cluster structure. In model-based clustering, these atypical observations are commonly referred to as outliers. The robust clustering literature distinguishes between mild outliers and gross outliers, as discussed by \cite{punzo2016parsimonious}. Mild outliers deviate from the assumed model but still originate from an underlying probabilistic mechanism \citep{ritter2014robust}. These cases often arise when the tails of the component distributions lack sufficient flexibility to accommodate extreme values. In such instances, the preferred strategy is to use more flexible cluster distributions that accommodate heavier tails and greater dispersion, rather than removing observations. In contrast, gross outliers are typically defined as observations that deviate from the data's primary generative structure \citep{ritter2014robust}. These observations are highly irregular and may result from mechanisms that are not easily modeled within the assumed distributional framework. Therefore, robust approaches often seek to mitigate their influence by trimming or suppressing \citep{cuesta1997trimmed}.

Recent robust approaches aim to model atypical observations within each cluster component, rather than excluding outliers from group membership. For instance, \cite{6042874} introduced a mixture of mixtures framework in which each cluster comprises a Gaussian component for structured observations and a uniform component for cluster-specific noise. This methodology facilitates the identification of diffuse observations while preserving the integrity of the underlying mixture structure. The uniform component assigns a constant density across its support, which prevents differentiation between moderately atypical observations and those that are extremely distant from the cluster center. As a result, all observations within the noise region receive equal support from the uniform component, regardless of their degree of extremeness. The absence of a distance-dependent decay mechanism may therefore limit the model's capacity to characterize the gradual transition from the central mass of a cluster to its most extreme observations.

An alternative component-wise approach is provided by the parsimonious mixtures of multivariate contaminated normal (CNM) distributions introduced by \cite{punzo2016parsimonious}. In this framework, each cluster is represented as a mixture of two Gaussian distributions: a primary reference distribution for typical observations and a secondary distribution with an inflated covariance matrix to account for atypical points.

Although both approaches provide effective mechanisms for robust clustering and outlier detection, they address atypical observations either through diffuse uniform noise regions or by inflating variance within the Gaussian family. Consequently, neither framework directly incorporates a heavy-tailed component distribution with a continuously decaying tail structure, which may limit their ability to capture data structures exhibiting heavy-tailed behavior. 

In the physical sciences, the Voigt profile, defined as a convolution of a Gaussian and a Cauchy (Lorentzian) distribution, is widely used to model such structures \citep{kielkopf1973voigt}. Despite its theoretical appeal, the Voigt profile is difficult to incorporate directly into likelihood-based mixture models because its density lacks a closed-form analytical expression, requiring computationally intensive numerical integration \citep{thompson1987voigt}. To address this challenge, practical implementations often use the pseudo-Voigt approximation, which represents the profile as a weighted linear combination of Gaussian and Cauchy components \citep{wertheim1974pseudo}. This approximation is computationally efficient and retains the essential heavy-tailed characteristics of the Voigt profile, making it suitable for complex statistical modeling.

To address the limitations of existing robust mixture models, we propose a multivariate pseudo-Voigt mixture model (MPVM) for clustering and outlier detection. In this framework, each cluster is represented as a two-component mixture of Gaussian and Cauchy distributions, offering a flexible distributional form that accommodates heavy-tailed behavior. A key feature of this formulation is that, regardless of which component dominates the central mass of the cluster, the Cauchy component, due to its polynomial decay, prevails in the tails of the distribution. This tail behavior offers a natural mechanism for identifying atypical observations within each cluster. Unlike contaminated normal models, which inflate the variance within a single distributional family, the MPVM framework utilizes the heavy-tailed nature of the Cauchy component to capture observations in the extreme regions of the data. Consequently, the proposed model enables simultaneous clustering and internal outlier detection.

The structure of this paper is as follows. Section~\ref{sec: background} provides the theoretical background and introduces the key concepts underlying the proposed framework. Section~\ref{sec: methodology} presents the proposed multivariate pseudo-Voigt mixture model, its parameter estimation procedure, and the mechanism for outlier detection. Section~\ref{sec: simulation} evaluates the performance of the proposed model through both simulation studies and applications to real data, comparing its clustering accuracy and outlier detection capabilities against established robust models. Finally, Section~\ref{sec: discussion} concludes with a summary of the findings and outlines directions for future research.

\section{Background}\label{sec: background}
Outliers are defined with respect to a reference distribution for regular observations and are typically considered as data points located in regions of low probability under that model \citep{DaviesGather1993,ritter2014robust}. However, the parameters of the reference distribution are generally unknown and must be estimated from data that may include atypical observations. This estimation process can result in masking, where outliers influence the estimated location or scatter, thereby diminishing their own apparent extremeness or that of other observations \citep{BeckerGather1999}. Mixture-based methods mitigate this problem by modeling regular and atypical observations as distinct components or subcomponents within a unified probabilistic framework.

\subsection{Finite Mixture Model}\label{subsec: finite mixture}
Let $\mathbf{x}_1, \ldots, \mathbf{x}_n$ be a sample of $n$ independent $p$-dimensional observations. We assume these data arise from a finite mixture of $G$ components with the marginal density
\begin{equation}
f(\mathbf{x}_i ; \boldsymbol{\Psi}) = \sum_{g=1}^G \pi_g f_g(\mathbf{x}_i ; \boldsymbol{\theta}_g),
\end{equation}
where $\pi_g > 0$ are the mixing proportions satisfying $\sum_{g=1}^G \pi_g = 1$, and $\boldsymbol{\Psi} = \{(\pi_g, \boldsymbol{\theta}_g)\}_{g=1}^G$ denotes the total parameter set, and \(f_g(\mathbf{x}_i ; \boldsymbol{\theta}_g) \) is the g$^{\textrm{th}}$  component density. While not strictly necessary, the component densities \(f_g(\mathbf{x}_i ; \boldsymbol{\theta}_g) \) are often assumed to belong to the same distributional family \citep{PeelMcLachlan2000, mcnicholas2016mixture}.

\subsection{Contaminated Normal Distribution}\label{subsec: contaminated normal}
The multivariate contaminated normal distribution, originally proposed by \cite{tukey1960} as a framework for studying robustness under contaminated sampling conditions, has subsequently been adapted for model-based clustering and outlier detection \citep{punzo2016parsimonious}. This distribution is defined as a convex combination of a reference normal component and a variance-inflated normal component, both sharing a common mean vector.

A $p$-dimensional random variable $\mathbf{X}$ is said to follow a multivariate contaminated normal distribution, denoted as $\mathbf{X} \sim \text{CN}_p(\boldsymbol{\mu}, \boldsymbol{\Sigma}, \alpha, \eta)$, if its probability density function is defined as follows:
\begin{equation}
f_{CN}(\mathbf{x}; \boldsymbol{\mu}, \boldsymbol{\Sigma}, \alpha, \eta) =  \alpha \phi_p(\mathbf{x}; \boldsymbol{\mu}, \boldsymbol{\Sigma}) + (1 - \alpha) \phi_p(\mathbf{x}; \boldsymbol{\mu}, \eta\boldsymbol{\Sigma}),
\end{equation}
Here, $\phi_p(\mathbf{x}; \boldsymbol{\mu}, \boldsymbol{\Sigma})$ represents the $p$-dimensional multivariate normal density with mean vector $\boldsymbol{\mu}$, covariance matrix $\boldsymbol{\Sigma}$, $\alpha_g$ is the proportion of good observations in cluster $g$, and $\eta_g > 1$ is the covariance inflation factor for the contaminated observations.

A finite mixture of contaminated normal distributions is defined as 
\begin{equation}
f(\mathbf{x}_i;\boldsymbol{\Psi})
=
\sum_{g=1}^{G} \pi_g f_{\mathrm{CN}} (\mathbf{x}_i; \boldsymbol{\mu}_g,
\boldsymbol{\Sigma}_g, \alpha_g,  \eta_g),
\label{eq:cn_mixture}
\end{equation}
where \(\pi_g\) denotes the mixing proportion of component \(g\), satisfying
\(\sum_{g=1}^{G}\pi_g = 1\), and \(f_{\mathrm{CN}}(\mathbf{x}_i; \boldsymbol{\mu}_g, \boldsymbol{\Sigma}_g, \alpha_g, \eta_g)\) denotes the multivariate contaminated normal density of component \(g\).

In a \(G\)-component contaminated normal mixture model with \(p\) dimensions, the free parameters include \(G-1\) mixing proportions, \(Gp\) mean parameters, \(Gp(p+1)/2\) covariance parameters, \(G\) contamination proportions \((\alpha_g)\), and \(G\) covariance inflation factors \((\eta_g)\). Therefore, the total number of free parameters is given by

\begin{equation}
q = (G-1) + Gp + \frac{Gp(p+1)}{2} + 2G.
\label{eq:cnm_num_parameters}
\end{equation}

Within this framework, the component dominating the central region corresponds to ``good'' observations, while the component with inflated covariance accounts for the tails and captures ``bad'' observations. To guarantee identifiability between the two normal components, the constraints $\alpha_g \geq 0.5$ and $\eta_g > 1$ are imposed. These conditions ensure that the majority of the probability mass is allocated to the ``good'' component with the smaller covariance, whereas the contaminated component, characterized by an inflated covariance matrix, models tail behavior and potential outliers.

While this formulation facilitates simultaneous clustering and outlier detection, it has limitations when applied to highly contaminated datasets or to data containing outliers from heavy-tailed distributions.

\subsection{Model-Based Clustering}\label{subsec: model-based clustering}
Let \(\mathbf{z}_i = (\mathbf{z}_{i1}, \ldots, \mathbf{z}_{iG})\) denote the component membership of observation \(i\), so that \(z_{ig} = 1\) if observation \(i\) belongs to component \(g\) and \(z_{ig} = 0\) otherwise. Within this framework \(\mathbf{z}_i\) is the realization of random variable \(\mathbf{Z}_i\), which follows a multinomial distribution with one draw over \(G\) categories with probabilities given by \( (\pi_1, \ldots, \pi_G)\). According to \cite{McLachlanPeel2000}, \(\mathbf{Z}_1, \ldots, \mathbf{Z}_n\) are assumed to to be independent and identically distributed. 
\[
\hat(z)_{ig}
=
\mathbb{E}[Z_{ig}\mid \mathbf{x}_i;\boldsymbol{\Psi}]
=
\Pr(Z_{ig}=1\mid \mathbf{x}_i;\boldsymbol{\Psi}).
\]
A mixing proportion \(\pi_g\) can be interpreted as the \emph{a priori} probability that observation \(x_i\) belongs to component \(g\). Hence, the corresponding \emph{a posteriori} probability is 
\begin{equation}
\hat(z)_{ig}
=
\frac{
\pi_g \, f_g(\mathbf{x}_i ; \boldsymbol{\theta}_g)
}{
\displaystyle
\sum_{h=1}^{G}
\pi_h \, f_h(\mathbf{x}_i ; \boldsymbol{\theta}_h)
}.
\label{eq: a posteriori tau}
\end{equation}
The MAP classification rule assigns observation $\mathbf{x}_i$ to the cluster
corresponding to the largest posterior probability, that is,
\begin{equation}
\widehat{z}_{ig}
=
\begin{cases}
1 & \text{if } 
g = \displaystyle\arg \max_{h \in \{1,\ldots,G\}} \hat(z)_{ih}, \\
0 & \text{otherwise}.
\end{cases}
\label{eq:MAP_rule}
\end{equation}

This rule produces a hard partition of the data into $G$ clusters, while the posterior probabilities $\hat(z)_{ig}$ themselves provide a natural measure of classification uncertainty.

\subsection{Convergence Criteria}\label{subsec: convergence criteria}
The EM algorithm generates a sequence of parameter estimates $\{\boldsymbol{\Psi}^{(r)}\}$ that monotonically increase the observed-data log-likelihood $\ell(\boldsymbol{\Psi})$. As described in \cite{aitken1926,McLachlanPeel2000}, convergence is evaluated using Aitken acceleration, which assumes that the log-likelihood increases form an approximately convergent geometric sequence near convergence. At iteration $r+1$, the acceleration factor is calculated as follows:
\[
a^{(r+1)}
=
\frac{
\ell(\boldsymbol{\Psi}^{(r+2)})
-
\ell(\boldsymbol{\Psi}^{(r+1)})
}{
\ell(\boldsymbol{\Psi}^{(r+1)})
-
\ell(\boldsymbol{\Psi}^{(r)})
}.
\]
This ratio provides an asymptotic estimate of the converged log-likelihood, given by
\[
\ell_{\infty}^{(r+2)}
=
\ell(\boldsymbol{\Psi}^{(r+1)})
+
\frac{
\ell(\boldsymbol{\Psi}^{(r+2)})
-
\ell(\boldsymbol{\Psi}^{(r+1)})
}{
1-a^{(r+1)}
}.
\]
Convergence is declared when the estimated remaining increase in the log-likelihood falls below a specified threshold, that is,
\[
\left|
\ell_{\infty}^{(r+2)}
-
\ell(\boldsymbol{\Psi}^{(r+1)})
\right|
<
\varepsilon,
\]
where $\varepsilon > 0$ denotes a predefined tolerance.

\subsection{Model Selection}\label{sec:model_selection}

In finite mixture modeling, determining the appropriate model complexity is essential to balance goodness of fit and parsimony. The number of clusters is a primary source of complexity, as increasing it typically improves fit to observed data but may lead to overfitting. To manage this trade-off, model selection criteria are employed to compare alternative models.

The Bayesian Information Criterion (BIC) \citep{Schwarz1978} is among the most widely used criteria. It integrates a measure of model fit with a penalty for model complexity. The BIC is defined as follows:

\begin{equation}
\mathrm{BIC}
=
-2\,\ell(\widehat{\boldsymbol{\Psi}})
+
q \log n,
\end{equation}

Here, $\ell(\widehat{\boldsymbol{\Psi}})$ represents the maximized observed-data log-likelihood, $n$ denotes the sample size, and $q$ is the number of free parameters in the model. Models with lower BIC values are preferred.

\section{Methodology}\label{sec3}
\label{sec: methodology}
\subsection{Mixture of Multivariate Pseudo-Voigt Distributions}
\label{subsec: 2.1 model design}
A multivariate pseudo-Voigt distribution is defined as a convex combination of multivariate Gaussian and multivariate Cauchy distributions that share a common location parameter. Its density can be written as
\begin{equation}
f_g(\mathbf{x}_i ; \boldsymbol{\theta}_g)
=
\alpha_g \,
\phi_p(\mathbf{x}_i ; \boldsymbol{\mu}_g, \boldsymbol{\Sigma}_g)
+
(1-\alpha_g)\,
C_p(\mathbf{x}_i ; \boldsymbol{\mu}_g, \boldsymbol{\Gamma}_g),
\label{eq:cluster_density}
\end{equation}
where $\alpha_g \in (0,1)$ is the within-cluster mixing proportion, 
$\boldsymbol{\mu}_g \in \mathbb{R}^p$ is the shared location parameter, 
$\boldsymbol{\Sigma}_g$ is the Gaussian covariance matrix, 
$\boldsymbol{\Gamma}_g$ is the Cauchy scale matrix, and \(\boldsymbol{\theta}_g
=
\{
\alpha_g,
\boldsymbol{\mu}_g,
\boldsymbol{\Sigma}_g,
\boldsymbol{\Gamma}_g
\} \) is the parameter vector for component $g$.

Let
\[
\delta(\mathbf{x}_i ; \boldsymbol{\mu}_g, \mathbf{A})
=
(\mathbf{x}_i - \boldsymbol{\mu}_g)^\top
\mathbf{A}^{-1}
(\mathbf{x}_i - \boldsymbol{\mu}_g)
\]
denote the squared Mahalanobis distance between $\mathbf{x}_i$ and
$\boldsymbol{\mu}_g$ with respect to a positive-definite matrix
$\mathbf{A}$. The multivariate Gaussian density is then given by
\begin{align}
\phi_p(\mathbf{x}_i ; \boldsymbol{\mu}_g, \boldsymbol{\Sigma}_g)
&=
(2\pi)^{-p/2}
|\boldsymbol{\Sigma}_g|^{-1/2}
\exp\left\{
-\frac{1}{2}
\delta(\mathbf{x}_i ; \boldsymbol{\mu}_g, \boldsymbol{\Sigma}_g)
\right\},
\end{align}
while the multivariate Cauchy density is
\begin{align}
C_p(\mathbf{x}_i ; \boldsymbol{\mu}_g, \boldsymbol{\Gamma}_g)
&=
\frac{
\Gamma\left(\frac{p+1}{2}\right)
}{
\pi^{(p+1)/2}
|\boldsymbol{\Gamma}_g|^{1/2}
\left[
1+
\delta(\mathbf{x}_i ; \boldsymbol{\mu}_g, \boldsymbol{\Gamma}_g)
\right]^{(p+1)/2}
}.
\end{align}

Following the finite mixture framework introduced in Section~\ref{subsec: finite mixture}, the mixture of multivariate pseudo-Voigt distributions is defined as
\begin{equation}
f(\mathbf{x}_i ; \boldsymbol{\Psi})
=
\sum_{g=1}^{G}
\pi_g
\Big[
\alpha_g \,
\phi_p(\mathbf{x}_i ; \boldsymbol{\mu}_g, \boldsymbol{\Sigma}_g)
\nonumber
+
(1-\alpha_g)\,
C_p(\mathbf{x}_i ; \boldsymbol{\mu}_g, \boldsymbol{\Gamma}_g)
\Big],
\label{eq:mpv_mixture}
\end{equation}
where $\pi_g > 0$ denotes the cluster mixing proportion and
$\sum_{g=1}^{G} \pi_g = 1$.

Although the pseudo-Voigt formulation permits the Gaussian and Cauchy components to have distinct scale matrices, the constraint
$\boldsymbol{\Gamma}_g = \boldsymbol{\Sigma}_g$
 is imposed to ensure that both components share a common geometric structure. In this framework, the Cauchy component introduces heavier tails while maintaining the overall orientation of the cluster. Beyond preserving a unified geometric structure, this constraint also reduces the number of free parameters to be estimated to
\begin{equation}
q = (G-1) + Gp + \frac{Gp(p+1)}{2} + G,
\label{eq:mpv_num_parameters}
\end{equation}
where \(G-1\) represents the mixing proportions \((\pi_g)\), \(Gp\) denotes the component mean vectors \((\boldsymbol{\mu}_g)\), \(p(p+1)/2\) corresponds to the component scale matrices \((\boldsymbol{\Sigma}_g)\), and \(G\) refers to the within-cluster Gaussian proportions \((\alpha_g)\). Compared with the contaminated normal mixture model, the proposed model requires \(G\) fewer parameters, resulting in a more parsimonious formulation.

\subsection{Parameter Estimation}\label{subsec: parameter estimation}
Parameter estimation is performed using an Expectation Maximization (EM) algorithm. Direct maximization of the observed-data log-likelihood is analytically challenging because of its log-sum structure. The EM algorithm instead iteratively maximizes the conditional expectation of the complete-data log-likelihood. 

\subsubsection{Latent Variable Structure}
\label{subsec:latent_structure}

Parameter estimation using the EM algorithm is facilitated by defining the complete-data structure with three sets of latent variables. The cluster membership indicator $Z_{ig}$ is defined such that $Z_{ig}=1$ if observation $\mathbf{x}_i$ belongs to cluster $g$ and $Z_{ig}=0$ otherwise, with $\sum_{g=1}^G Z_{ig} = 1$. Given $Z_{ig}=1$, a within-cluster component indicator $V_{ig}$ is introduced, where $V_{ig}=1$ if $\mathbf{x}_i$ originates from the Gaussian component and $v_{ig}=0$ if it originates from the Cauchy component. The probabilities are specified as $\Pr(V_{ig}=1 \mid Z_{ig}=1) = \alpha_g$ and $\Pr(V_{ig}=0 \mid Z_{ig}=1) = 1-\alpha_g$.

While the latent variables $Z_{ig}$ and $V_{ig}$ provides the complete data-likelihood, the EM algorithm does not provide a closed-form parameter estimates for $\mathbf{\mu}$ and $\mathbf{\Sigma}$. This challenge is addressed by utilizing the representation of the multivariate Cauchy distribution as a scale mixture of the Gaussian distribution \citep{PeelMcLachlan2000}.

For observations arising from the Cauchy component ($V_{ig}=0$), an additional latent scaling variable $U_{ig}$ is introduced such that
\begin{equation}
\begin{aligned}
U_{ig} 
&\mid (Z_{ig}=1, V_{ig}=0)
&&\sim \mathrm{Gamma}\left(\frac{1}{2}, \frac{1}{2}\right), \\
\mathbf{X}_i
&\mid (U_{ig}=u_{ig}, Z_{ig}=1, V_{ig}=0)
&&\sim \mathcal{N}_p\left(
\boldsymbol{\mu}_g,
\frac{\boldsymbol{\Sigma}_g}{u_{ig}}
\right).
\end{aligned}
\label{eq:scale_mix}
\end{equation}

\noindent Hence, the marginal distribution of $\mathbf{X}_i$ will express the Cauchy density as
\[
C_p(\mathbf{x}_i ; \boldsymbol{\mu}_g, \boldsymbol{\Sigma}_g)
=
\int_0^\infty
h(u_{ig})
\,
\phi_p
\left(
\mathbf{x}_i ;
\boldsymbol{\mu}_g,
\frac{\boldsymbol{\Sigma}_g}{u_{ig}}
\right)
\,du_{ig},
\]
where $h(u_{ig})$ denotes the Gamma density with shape parameter $1/2$ and rate parameter $1/2$. Therefore, conditional on $\boldsymbol{U}_{ig} = u_{ig}$, the Cauchy contribution is represented through the joint density
\[
p(\mathbf{x}_i,u_{ig})
=
h(u_{ig})
\,
\phi_p
\left(
\mathbf{x}_i ;
\boldsymbol{\mu}_g,
\frac{\boldsymbol{\Sigma}_g}{u_{ig}}
\right).
\]
This formulation enables closed-form parameter estimation within the EM framework.

\subsubsection{Complete Data Log-likelihood}\label{subsubsec:complete data-loglikelihood}
Introducing the latent indicator vectors $\boldsymbol{Z}_i$ and
$\boldsymbol{V}_i$, the complete-data density for observation
$\boldsymbol{x}_i$ is given by
\begin{equation}
\begin{aligned}
P(\mathbf{z}_i,\mathbf{v}_i,\mathbf{x}_i \mid \boldsymbol{\Psi})
=
\prod_{g=1}^G
\Big[
&
\pi_g
\left\{
\alpha_g
\phi_p(\mathbf{x}_i \mid \boldsymbol{\mu}_g,\boldsymbol{\Sigma}_g)
\right\}^{v_{ig}}
\\
&
\times
\left\{
(1-\alpha_g)
C_p(\mathbf{x}_i \mid \boldsymbol{\mu}_g,\boldsymbol{\Sigma}_g)
\right\}^{1-v_{ig}}
\Big]^{z_{ig}}.
\end{aligned}
\label{eq:joint_density}
\end{equation}

Using the scale-mixture representation introduced in the section \ref{subsec:latent_structure}, the Cauchy density can be expressed through the latent scaling variable $U_{ig}$. Let $\boldsymbol{\Xi}_i (\boldsymbol{Z}_i,\boldsymbol{V}_i,\boldsymbol{U}_i)$ denote the collection of latent variables associated with observation $\boldsymbol{x}_i$. The complete-data log-likelihood,
$\ell_c(\boldsymbol{\Psi}\mid \boldsymbol{X},\boldsymbol{\Xi})$, becomes
\begin{equation}
\begin{aligned}
\ell_c(\boldsymbol{\Psi}\mid \boldsymbol{X},\boldsymbol{\Xi})
=
\sum_{i=1}^n
\sum_{g=1}^G
z_{ig}
\Biggl(
&
\log \pi_g
+
v_{ig}\log \alpha_g
+
(1-v_{ig})\log(1-\alpha_g)
\\
&
+
v_{ig}
\log
\phi_p
(
\mathbf{x}_i ;
\boldsymbol{\mu}_g,
\boldsymbol{\Sigma}_g
)
\\
&
+
(1-v_{ig})
\log
\phi_p
\left(
\mathbf{x}_i ;
\boldsymbol{\mu}_g,
\frac{\boldsymbol{\Sigma}_g}{u_{ig}}
\right)
\\
&
+
(1-v_{ig})
\log h(u_{ig})
\Biggr).
\end{aligned}
\label{eq:complete_data_loglik}
\end{equation}

\subsubsection{E-step}
\label{subsubsec:estep}

At iteration $r+1$, the E-step computes the conditional expectations of the latent variables given the observed data and the current parameter estimates $\boldsymbol{\Psi}^{(r)}$.

The posterior probability that observation $\mathbf{x}_i$ belongs to
cluster $g$ at iteration $r$ is given by
\begin{equation}
\hat{z}_{ig}^{(r)}
=
\mathbb{E}[Z_{ig}\mid \mathbf{x}_i;\boldsymbol{\Psi}^{(r)}]
=
\frac{
\pi_g^{(r)}
f_g(\mathbf{x}_i;\boldsymbol{\theta}_g^{(r)})
}{
\displaystyle
\sum_{h=1}^{G}
\pi_h^{(r)}
f_h(\mathbf{x}_i;\boldsymbol{\theta}_h^{(r)})
}.
\label{eq:tau_update}
\end{equation}

Conditional on membership in cluster $g$, the posterior probability that observation $\mathbf{x}_i$ arises from the Gaussian within-cluster component is given by
\begin{equation}
\hat{v}_{ig}^{(r)}
=
\mathbb{E}[V_{ig} \mid \mathbf{x}_i, Z_{ig}=1; \boldsymbol{\Psi}^{(r)}]
=
\frac{
\alpha_g^{(r)}
\phi_p(\mathbf{x}_i ; \boldsymbol{\mu}_g^{(r)}, \boldsymbol{\Sigma}_g^{(r)})
}{
f_g(\mathbf{x}_i ; \boldsymbol{\theta}_g^{(r)})
}.
\label{eq:w_paper}
\end{equation}
Accordingly, the posterior probability that $\mathbf{x}_i$ belongs to
the Cauchy within-cluster component is given by
\[
1-\hat{v}_{ig}^{(r)}.
\]

The conditional expectation of the latent scaling variable is given by
\begin{equation}
u_{ig}^{(r)}
=
\mathbb{E}[U_{ig}\mid \mathbf{x}_i,Z_{ig}=1,V_{ig}=0;\boldsymbol{\Psi}^{(r)}].
\end{equation}

\noindent By Bayes' theorem, the conditional density of $U_{ig}$ is proportional to
\[
p(u_{ig}\mid \mathbf{x}_i,Z_{ig}=1,V_{ig}=0)
\propto
u_{ig}^{(p+1)/2-1}
\exp\left\{
-\frac{1+\delta_{ig}^{(r)}}{2}u_{ig}
\right\},
\]
where
\[
\delta_{ig}^{(r)}
=
(\mathbf{x}_i-\boldsymbol{\mu}_g^{(r)})^\top
(\boldsymbol{\Sigma}_g^{(r)})^{-1}
(\mathbf{x}_i-\boldsymbol{\mu}_g^{(r)}).
\]
Therefore,
\[
U_{ig}\mid(\mathbf{x}_i,Z_{ig}=1,V_{ig}=0;\boldsymbol{\Psi}^{(r)})
\sim
\mathrm{Gamma}
\left(
\frac{p+1}{2},
\frac{1+\delta_{ig}^{(r)}}{2}
\right),
\]
and the conditional expectation required in the E-step becomes
\begin{equation}
u_{ig}^{(r)}
=
\mathbb{E}[U_{ig}\mid \mathbf{x}_i,Z_{ig}=1,V_{ig}=0;\boldsymbol{\Psi}^{(r)}]
=
\frac{p+1}{1+\delta_{ig}^{(r)}}.
\label{eq:u_ig}
\end{equation}

\subsubsection{M-steps}
\label{subsubsec:cmsteps}
The EM algorithm proceeds by iteratively maximizing the conditional expectation of the complete-data log-likelihood,
\[
Q(\boldsymbol{\Psi};\boldsymbol{\Psi}^{(r)})
=
\mathbb{E}_{\boldsymbol{\Xi}\mid\boldsymbol{X};\boldsymbol{\Psi}^{(r)}}
\left[
\ell_c(\boldsymbol{\Psi}\mid\boldsymbol{X},\boldsymbol{\Xi})
\right].
\]
with respect to subsets of parameters while holding the remaining parameters fixed, in a sequential manner. For parameter estimation, it is convenient to partition
$\ell_c(\boldsymbol{\Psi}\mid\boldsymbol{X},\boldsymbol{\Xi})$ into additive blocks as
\begin{equation}
\ell_c(\boldsymbol{\Psi}\mid\boldsymbol{X},\boldsymbol{\Xi})
=
\ell_{1c}(\boldsymbol{\pi})
+
\ell_{2c}(\boldsymbol{\alpha})
+
\ell_{3c}(\boldsymbol{\mu},\boldsymbol{\Sigma}),
\label{eq:lc_partition_mpv}
\end{equation}
where
\begin{align}
\ell_{1c}(\boldsymbol{\pi})
&=
\sum_{i=1}^n
\sum_{g=1}^G
z_{ig}\log\pi_g,
\label{eq:l1c_pi}
\\[4pt]
\ell_{2c}(\boldsymbol{\alpha})
&=
\sum_{i=1}^n
\sum_{g=1}^G
z_{ig}
\Big[
v_{ig}\log\alpha_g
+
(1-v_{ig})\log(1-\alpha_g)
\Big],
\label{eq:l2c_alpha}
\end{align}
and the remaining block contains the terms involving the shared location and scale parameters:
\begin{align}
\ell_{3c}(\boldsymbol{\mu},\boldsymbol{\Sigma})
&=
-\frac{1}{2}
\sum_{i=1}^n
\sum_{g=1}^G
z_{ig}v_{ig}
\left[
\log|\boldsymbol{\Sigma}_g|
+
\delta(\mathbf{x}_i;\boldsymbol{\mu}_g,\boldsymbol{\Sigma}_g)
\right]
\notag\\
&\quad
-\frac{1}{2}
\sum_{i=1}^n
\sum_{g=1}^G
z_{ig}(1-v_{ig})
\left[
\log|\boldsymbol{\Sigma}_g|
+
u_{ig}\,
\delta(\mathbf{x}_i;\boldsymbol{\mu}_g,\boldsymbol{\Sigma}_g)
\right]
+\text{const}.
\label{eq:l3c_simplified}
\end{align}

The mixing proportions and within-cluster Gaussian proportions can be updated using the following closed-form expressions:
\begin{equation}
\pi_g^{(r+1)} = \frac{1}{n} \sum_{i=1}^n \hat{z}_{ig}^{(r)} \quad \text{and} \quad \alpha_g^{(r+1)} = \frac{\sum_{i=1}^n \hat{z}_{ig}^{(r)} \hat{v}_{ig}^{(r)}}{n_g^{(r)}},
\label{eq:pi_alpha_update}
\end{equation}
where $n_g^{(r)} = \sum_{i=1}^n \hat{z}_{ig}^{(r)}$ denotes the effective size of cluster $g$.

The update for the location parameter is
\begin{equation}
\boldsymbol{\mu}_g^{(r+1)} = \frac{\sum_{i=1}^n \hat{z}_{ig}^{(r)} \left[ \hat{v}_{ig}^{(r)} + (1-\hat{v}_{ig}^{(r)}) \hat{u}_{ig}^{(r)} \right] \mathbf{x}_i}{\sum_{i=1}^n \hat{z}_{ig}^{(r)} \left[ \hat{v}_{ig}^{(r)} + (1-\hat{v}_{ig}^{(r)}) \hat{u}_{ig}^{(r)} \right]},
\label{eq:mu_update_unified}
\end{equation}

And, maximizing with respect to the shared covariance matrix $\boldsymbol{\Sigma}_g$ gives the following update:
\begin{equation}
\boldsymbol{\Sigma}_g^{(r+1)} = \frac{1}{n_g^{(r)}} \sum_{i=1}^n \hat{z}_{ig}^{(r)} \left[ \hat{v}_{ig}^{(r)} + (1-\hat{v}_{ig}^{(r)}) \hat{u}_{ig}^{(r)} \right] (\mathbf{x}_i - \boldsymbol{\mu}_g^{(r+1)})(\mathbf{x}_i - \boldsymbol{\mu}_g^{(r+1)})^\top.
\label{eq:Sigma_update_unified}
\end{equation}

\subsection{Outlier Detection}
\label{subsect : outlier_detect}

Outlier detection in mixture models frequently relies on latent sub-cluster membership. In contaminated mixture models, observations assigned to the contamination component are generally classified as outliers, whereas those assigned to the main component are considered typical. This interpretation is supported by the model's structure, which ensures that the primary component dominates the central region of the cluster and the contamination component dominates the tails. For instance, contaminated normal mixture models impose constraints such as $\alpha_g > 0.5$ and $\eta_g > 1$, which ensure that the normal component with covariance $\Sigma_g$ represents the cluster core, while the inflated normal component with covariance $\eta_g \Sigma_g$ captures more dispersed observations.

Although these constraints provide a clear interpretation of the contamination component
, they also limit the diversity of cluster structures that can be modeled. Specifically, they assume that atypical observations form a minority within the cluster and that contamination can be adequately described by a variance-inflated version of the main distribution. These assumptions become restrictive when the cluster displays genuinely heavy-tailed behavior or when tail-associated observations comprise a significant portion of the cluster mass.

The proposed multivariate pseudo-Voigt mixture model represents a departure from this framework. Since no restrictions are imposed on the within-cluster Gaussian proportion $\alpha_g$, the Cauchy sub-component is not limited to representing only tail observations. Depending on the estimated parameters, the Cauchy sub-component may contribute significantly to the cluster center and may even dominate the Gaussian sub-component near the common location parameter $\boldsymbol{\mu}_g$. Therefore, latent sub-cluster membership alone is insufficient for defining an observation as an outlier. An observation may have a high probability of originating from the Cauchy sub-component while still occupying a central region of the cluster.

Accordingly, outlier detection in the proposed framework relies on two complementary forms of evidence. The first is a generative criterion, which evaluates the posterior probability that an observation originates from the Cauchy sub-component. The second is a geometric criterion, which assesses whether the observation lies sufficiently far into the tail region relative to the cluster center. An observation is classified as an outlier only when both conditions are satisfied.

\subsubsection{The Ratio-at-Center Threshold}

To formalize this logic, we define the relative density ratio $R(\mathbf{x})$, which compares the weighted Cauchy sub-component to the weighted Gaussian sub-component at \(\mathbf{x}\):
\begin{equation}
R(\mathbf{x}) = \frac{(1-\alpha_g) C_p(\mathbf{x} \mid \boldsymbol{\mu}_g, \boldsymbol{\Sigma}_g)}{\alpha_g \phi_p(\mathbf{x} \mid \boldsymbol{\mu}_g, \boldsymbol{\Sigma}_g)}.
\label{eq:ratio_definition}
\end{equation}
The value of this ratio at the cluster center, $R(\boldsymbol{\mu}_g)$, serves as the critical reference threshold. This threshold represents the `maximum acceptable' relative dominance of the Cauchy sub-component for an observation to still be considered typical of the cluster's core structure.

Appendix~\ref{secA1} demonstrates that the relative log-ratio
\[
r_g(\mathbf{x})
=
\log\left\{
\frac{R(\mathbf{x})}{R(\boldsymbol{\mu}_g)}
\right\}
\]
depends only on the squared Mahalanobis distance from the cluster center.
\[r_g(\mathbf{x}) = \frac{1}{2}\delta(\mathbf{x}) - \frac{p+1}{2} \log\left(1+\delta(\mathbf{x})\right).
\]
It follows that there exists a unique dominance threshold, denoted as $\delta_g^\star$, such that observations satisfying
\[
\delta_g(\mathbf{x})>\delta_g^\star
\]
are located in a region where the Cauchy component dominates the Gaussian component. As a result, the boundary between the central and tail regions is defined by a single ellipsoidal contour. Beyond this threshold, the normalized log-ratio increases monotonically with Mahalanobis distance, ensuring that observations at greater distances remain within the tail region. This framework offers a principled, data-driven criterion for outlier detection, with the parameters $\alpha_g$ and $\boldsymbol{\Sigma}_g$ specifying the acceptable level of Cauchy dominance at the cluster center.

\subsubsection{Outlier Identification Criterion}

An observation $\mathbf{x}_i$ is formally classified as an outlier if it satisfies two distinct criteria representing generative and geometric evidence:

\begin{enumerate}
    \item \textbf{Generative Evidence:} The observation is more likely to have arisen from the Cauchy (contaminated) sub-component than the Gaussian sub-component:
    \begin{equation}
    \hat{v}_{ig} \le 0.5.
    \end{equation}
    
    \item \textbf{Geometric Evidence:} The observation's relative density ratio exceeds the baseline ratio established at the cluster center:
    \begin{equation}
    R(\mathbf{x}_i) > R(\boldsymbol{\mu}_g).
    \end{equation}
\end{enumerate}

A unique dominance threshold outlines the boundary between the central and tail regions of each cluster. Within the proposed dual-criterion framework, only observations originating from the tail region of their respective cluster are classified as outliers. As a result, central observations are not considered outliers, even if the Cauchy component is predominant at the cluster peak.

\begin{figure}[htbp]
    \centering

    \begin{subfigure}{0.48\textwidth}
        \centering
        \includegraphics[width=\linewidth]{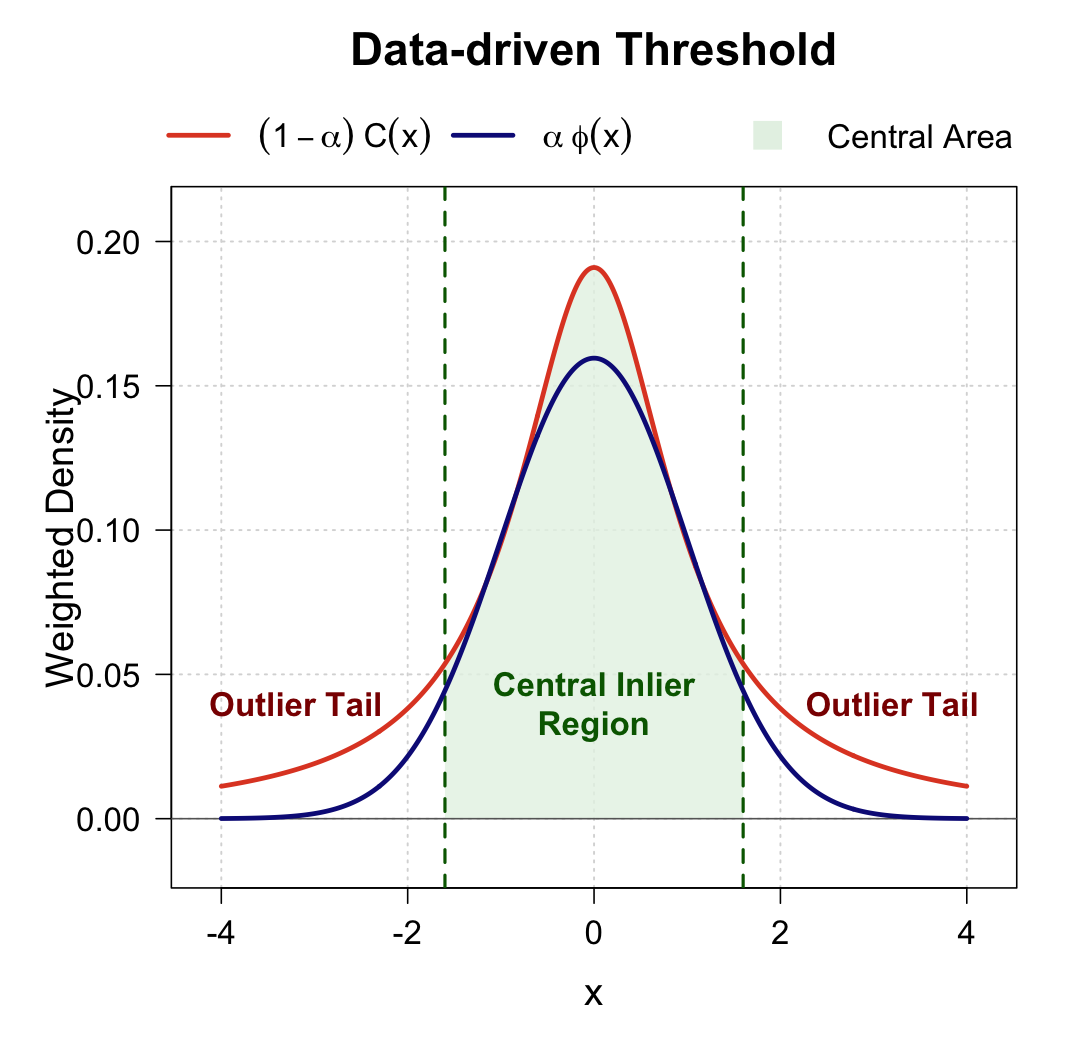}
        \caption{Weighted Densities}
        \label{fig:density}
    \end{subfigure}
    \hfill
    \begin{subfigure}{0.48\textwidth}
        \centering
        \includegraphics[width=\linewidth]{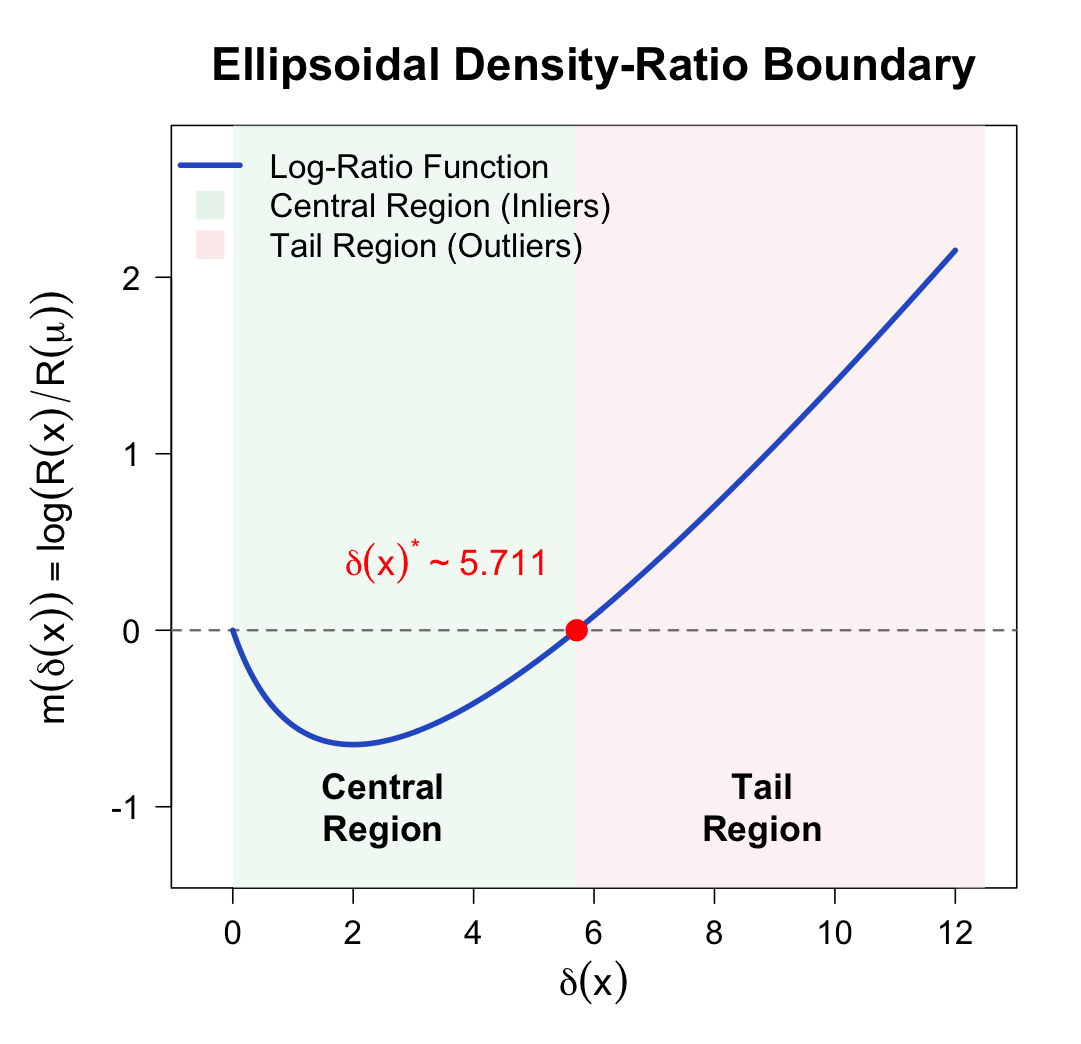}
        \caption{Relative Dominance Boundary}
        \label{fig:logratio}
    \end{subfigure}
   \caption{The Dual-Evidence Mechanism for Outlier Detection.
    \textbf{(a) Weighted Density Comparison:} An illustration of the sub-component densities with $\alpha = 0.4$. 
    \textbf{(b) Relative Log-Ratio Function:} The function $\log\{R(\mathbf{x})/R(\boldsymbol{\mu})\}$ with $p = 2$ dimensions plotted as a function of the squared Mahalanobis distance $\delta(x)$. The figure illustrates the existence of a unique positive root $\delta^\star$, at which the relative density ratio returns to its baseline value at the cluster center. Observations satisfying $\delta(\mathbf{x}) > \delta^\star$ lie in the tail region.}
    \label{fig:ratio_combined}
\end{figure}
The proposed framework provides two primary advantages over traditional robust mixture models. First, by integrating a Gaussian core with a heavy-tailed Cauchy component that shares a common mean $\boldsymbol{\mu}$, the model captures extreme observations while preserving a stable estimation of the cluster center. This structure enables separate modeling of the core and tail behavior within a cluster, thereby reducing the influence of extreme observations on location estimation. Second, the model removes the need for structural constraints that require the clean component to dominate the cluster, such as $\alpha > 0.5$. Consequently, the framework is capable of accommodating clusters that exhibit significant heavy-tailed behavior or contamination levels that constitute a large proportion of the cluster mass.

\subsection{Initialization}
\label{subsec:initialization}

The initialization strategy utilizes a robust partition derived from trimmed $k$-means. The set $\mathcal{K}$ represents observations retained after trimming, while $\mathcal{T}$ denotes trimmed observations. Observations in $\mathcal{K}$ receive hard cluster memberships according to the trimmed $k$-means partition and are initialized in the Gaussian sub-component. Observations in $\mathcal{T}$ receive soft cluster memberships, determined by normalized inverse Euclidean distances to the cluster centers, and are initialized in the Cauchy sub-component. These latent memberships and sub-component indicators establish the initial parameter configuration for the EM algorithm. The procedure is outlined in Algorithm~\ref{alg:mpv_init}.
\begin{algorithm}[htbp]
\caption{MPVM Initialization Procedure}
\label{alg:mpv_init}
\small
\textbf{Input:} $\mathbf{X}$, $G$, $\alpha_{\mathrm{trim}}$
\begin{algorithmic}[1]
\State Apply trimmed $k$-means
\State Partition observations into:
\Statex \hspace{1em} $\mathcal{K}$: retained observations
\Statex \hspace{1em} $\mathcal{T}$: trimmed observations
\For{$\mathbf{x}_i \in \mathcal{K}$}
    \State $Z_{ig}^{(0)} \leftarrow$ hard assignment from trimmed $k$-means
    \State $V_{ig}^{(0)} \leftarrow 1$
\EndFor
\For{$\mathbf{x}_i \in \mathcal{T}$}
    \State $Z_{ig}^{(0)} \leftarrow$ soft assignment based on normalized inverse Euclidean distances
    \State $V_{ig}^{(0)} \leftarrow 0$
\EndFor
\State \textbf{Return:} $\mathbf{Z}^{(0)}, \mathbf{V}^{(0)}$
\end{algorithmic}
\end{algorithm}

\section{Simulation Studies and Real Data Analysis}
\label{sec: simulation}

The performance of the proposed multivariate pseudo-Voigt mixture model is assessed using both simulation studies and applications to real datasets. This investigation focuses on two primary objectives: first, assessing the model's capacity to recover the underlying group structure and select the correct number of mixture components; and second, examining the model's robustness in identifying atypical observations within each cluster. These objectives are investigated across both controlled simulation environments and real-world datasets.

The analyses are organized into three distinct parts:
\begin{enumerate}
    \item \textbf{Edge-Case Analysis:} Controlled scenarios involving one and two clusters are first considered to investigate the fundamental behavior of the proposed model under boundary conditions.
    
    \item \textbf{Cluster Recovery and Outlier Detection:} The proposed model is evaluated using datasets generated from known cluster structures with varying levels of artificial contamination. In both the edge-case and recovery studies, the proposed model is benchmarked against mixtures of contaminated normal (CNM) distributions.
    
    \item \textbf{Real Data Application:} Finally, the proposed model is applied to benchmark datasets to assess goodness-of-fit and clustering performance. In this stage, comparisons are extended to include GMM, mixtures of multivariate $t$-distributions, and CNM models.
\end{enumerate}

\subsection{Computational Setup and Tools}
\label{subsect:computational_setup}

All computational routines for the proposed framework were implemented in the \textsf{R} statistical computing environment \citep{RCoreTeam2025}. For numerical stability during parameter estimation, Cholesky and eigenvalue decompositions were computed using corresponding built-in \textsf{R} functions. 

The robust initialization strategy outlined in Section~\ref{subsec:initialization} was executed using the \texttt{tkmeans()} function from the \texttt{tclust} package \citep{FritzGarcíaEscuderoMayo2012}. All simulations were conducted on a macOS Tahoe 26.3.1 system equipped with an Apple M3 Pro processor and 18 GB of RAM.

\subsection{Initialization Study}
\label{sub:init_study}

Initialization is critical in mixture model estimation because different starting values can lead to convergence to distinct local maxima of the likelihood function \citep{McLachlanPeel2000, Biernacki2003}. Although the \texttt{CNM} package \citep{Punzo2018} provides several initialization strategies, including GMM and $k$-means, the proposed multivariate pseudo-Voigt model requires an approach that accommodates the data's heavy-tailed characteristics. Therefore, an initialization procedure based on a modified trimmed $k$-means algorithm was developed.

A unified initialization framework was implemented to facilitate fair comparisons across all simulation experiments. This approach ensures that observed performance differences are attributable to the model architecture rather than to model-specific starting values. To assess the impact of initialization on model estimation, a preliminary experiment compared three strategies: random initialization, GMM-based initialization, and the proposed modified trimmed $k$-means approach. The findings from this comparison guided the selection of the initialization protocol for all subsequent simulation and real-data analyses.

\subsubsection{Random Initialization}
The random initialization strategy generates cluster membership indicators $Z_{ig}$ and within-cluster component membership indicators $V_{ig}$ stochastically. To mitigate the risk of poor local optima, ten independent random configurations are generated. For each configuration, the EM algorithm is run for five iterations. The configuration with the highest log-likelihood is then selected as the starting point for the full estimation procedure.

\subsubsection{GMM and Trimmed $k$-means Initialization}
Initialization via a GMM utilizes the \texttt{mclust} package \citep{Scrucca2016} to determine the initial cluster memberships $Z_{ig}$. The component membership indicators $V_{ig}$ are then assigned by designating a specified percentile of observations closest to the cluster center $\boldsymbol{\mu}_g$ as the Gaussian core ($V_{ig}=1$). The modified trimmed $k$-means initialization follows the methodology detailed in Section~\ref{subsec:initialization}, leveraging robust partitioning to diminish the influence of extreme observations prior to the first iteration of the EM algorithm.

\subsubsection{Simulation Design for the Initialization Strategy}

To evaluate the proposed initialization strategy, synthetic datasets were generated from a multivariate pseudo-Voigt mixture distributions with $G=3$ clusters in $p=2$ dimensions. The total sample size was set to $n=600$ with mixing proportions $\boldsymbol{\pi} = (0.2, 0.3, 0.5)$ and cluster-specific Gaussian proportions $\boldsymbol{\alpha} = (0.70, 0.60, 0.65)$. Two experimental settings were considered:
\begin{enumerate}
    \item Moderate-overlap clusters.
    \item Well-separated clusters.
\end{enumerate}

In the moderate-overlap setting, cluster locations were defined as $\boldsymbol{\mu}_1 = (0,0)$, $\boldsymbol{\mu}_2 = (0,-3)$, and $\boldsymbol{\mu}_3 = (3,0)$. In the well-separated setting, these were increased to $\boldsymbol{\mu}_1 = (0,0)$, $\boldsymbol{\mu}_2 = (0,-6)$, and $\boldsymbol{\mu}_3 = (6,0)$. For each cluster, the Gaussian and Cauchy components share a common scale structure, defined as:
\[
\boldsymbol{\Sigma}_1 = 
\begin{pmatrix} 1 & -0.5 \\ -0.5 & 1 \end{pmatrix}, \quad 
\boldsymbol{\Sigma}_2 = 
\begin{pmatrix} 1 & 0.5 \\ 0.5 & 1 \end{pmatrix}, \quad 
\boldsymbol{\Sigma}_3 = 
\begin{pmatrix} 1 & 0 \\ 0 & 1 \end{pmatrix}.
\]



For each dataset, both the proposed MPVM and the CNM  were fitted using three initialization strategies: random, GMM-based, and the modified trimmed $k$-means approach.

\subsubsection{Results and Discussion}

The performance of the initialization strategies was assessed using pairwise differences in the BIC relative to the modified trimmed $k$-means method. The mean and standard deviation (SD) of these differences ($\Delta \mathrm{BIC} = \mathrm{BIC}_{\text{Other}} - \mathrm{BIC}_{\text{Trimmed}}$) over 58 trials in the first scenario and 82 trials in the second scenario are summarized in Tables~\ref{tab:init_moderate} and \ref{tab:init_well}. Positive values indicate that the modified trimmed $k$-means initialization achieved a lower (superior) BIC.


\begin{table}[htbp]
\centering
\caption{Average pairwise BIC differences relative to the modified trimmed $k$-means initialization (Moderate Separation).}
\label{tab:init_moderate}
\setlength{\tabcolsep}{14pt}
\begin{tabular}{l r r r r}
\toprule
& \multicolumn{2}{c}{MPV Model} & \multicolumn{2}{c}{CNM Model} \\
\cmidrule(r){2-3} \cmidrule(l){4-5}
Comparison & Mean $\Delta \mathrm{BIC}$ & SD & Mean $\Delta \mathrm{BIC}$ & SD \\
\midrule
Random $-$ T $k$-means & 14.51 & 23.36 & 15.77 & 32.52 \\
GMM $-$ T $k$-means    & 32.19 & 36.60 & 21.49 & 31.39 \\
\bottomrule
\end{tabular}
\end{table}

\begin{table}[htbp]
\centering
\caption{Average pairwise BIC differences relative to the modified trimmed $k$-means initialization(Well-Separated Clusters).}
\label{tab:init_well}
\setlength{\tabcolsep}{15pt}
\begin{tabular}{l r r r r}
\toprule
& \multicolumn{2}{c}{MPV Model} & \multicolumn{2}{c}{CNM Model} \\
\cmidrule(r){2-3} \cmidrule(l){4-5}
Comparison & Mean $\Delta \mathrm{BIC}$ & SD & Mean $\Delta \mathrm{BIC}$ & SD \\
\midrule
Random $-$ T $k$-means & 14.24 & 19.84 & 8.39 & 96.10 \\
GMM $-$ T $k$-means    & 32.82 & 29.43 & 20.38 & 57.53 \\
\bottomrule
\end{tabular}
\end{table}

The results suggest that, for the proposed MPVM model, the modified trimmed k-means initialization generally provides better starting values than the random and GMM-based alternatives across the separation scenarios considered. The positive mean $\Delta\mathrm{BIC}$ values indicate that the trimmed approach tends to produce more stable initial partitions, reducing the influence of heavy-tailed observations during the early stages of estimation.

A similar trend is observed for the CNM model, particularly in the well-separated case. Consequently, the modified trimmed $k$-means strategy is adopted as the standard initialization for all subsequent analyses.

\subsection{Behavioural Study Under Edge Cases}

A behavioural study was conducted to investigate the properties of the proposed model under a set of edge-case scenarios. The analysis examines the behaviour of competing mixture models in extreme settings where the underlying data structure approaches the boundaries of the model formulation. 


For each scenario, a single-component sample of size $n = 1000$ observations was generated  in $p = 2 $ dimensions with 
\[
\boldsymbol{\mu}_1 = (0,0), \quad
\eta = 3, \quad
\boldsymbol{\Sigma}_1 = 
\begin{pmatrix}
1 & -0.5 \\
-0.5 & 1
\end{pmatrix}
.
\]

\noindent The following scenarios were considered:
\begin{enumerate}[leftmargin=1 cm ]
    \item CNM ($\alpha=1$), corresponding to a multivariate normal distribution.
    \item MPVM ($\alpha=0.5$).
    \item CNM ($\alpha=0.5$).
    \item MPVM ($\alpha=0$), corresponding to a multivariate Cauchy distribution.
\end{enumerate}

The results illustrate how the two models represent data structures under extreme conditions and identify outliers accordingly.

In Case 1, where the data follow a multivariate normal distribution, both models correctly identify a single cluster (G=1) and produce nearly identical estimates of the cluster center (Table~\ref{tab:case1}). Neither model classifies any observations as outliers. While both models represent the same underlying data structure, the MPVM model requires one fewer free parameter than the CNM (Equation~\eqref{eq:mpv_num_parameters} versus Equation~\eqref{eq:cnm_num_parameters}). Consequently, under the BIC criterion, MPVM incurs a smaller complexity penalty, leading to an expected reduction of approximately $\log(1000)\approx 6.91$ BIC units.

\begin{table}[htbp]
\centering
\caption{Summary of model estimates and information criteria for case 1.}
\label{tab:case1}
\setlength{\tabcolsep}{11pt}
\begin{tabular}{llccc p{2.8 cm}}
\toprule
Scenario & Model & $\hat{G}$ & BIC & \# Free Par & Estimated $\hat{\mu}$ \\
\midrule
\textbf{Case 1} & MPVM & 1 & 5472.8 & 6& $(0.001,-0.002)$ \\
& CNM & 1 & 5479.7 & 7& $(0.001,-0.002)$ \\

\bottomrule
\end{tabular}
\end{table}

\begin{figure}[htbp]
\centering

\begin{subfigure}{0.325\textwidth}
    \centering
    \includegraphics[width=\linewidth]{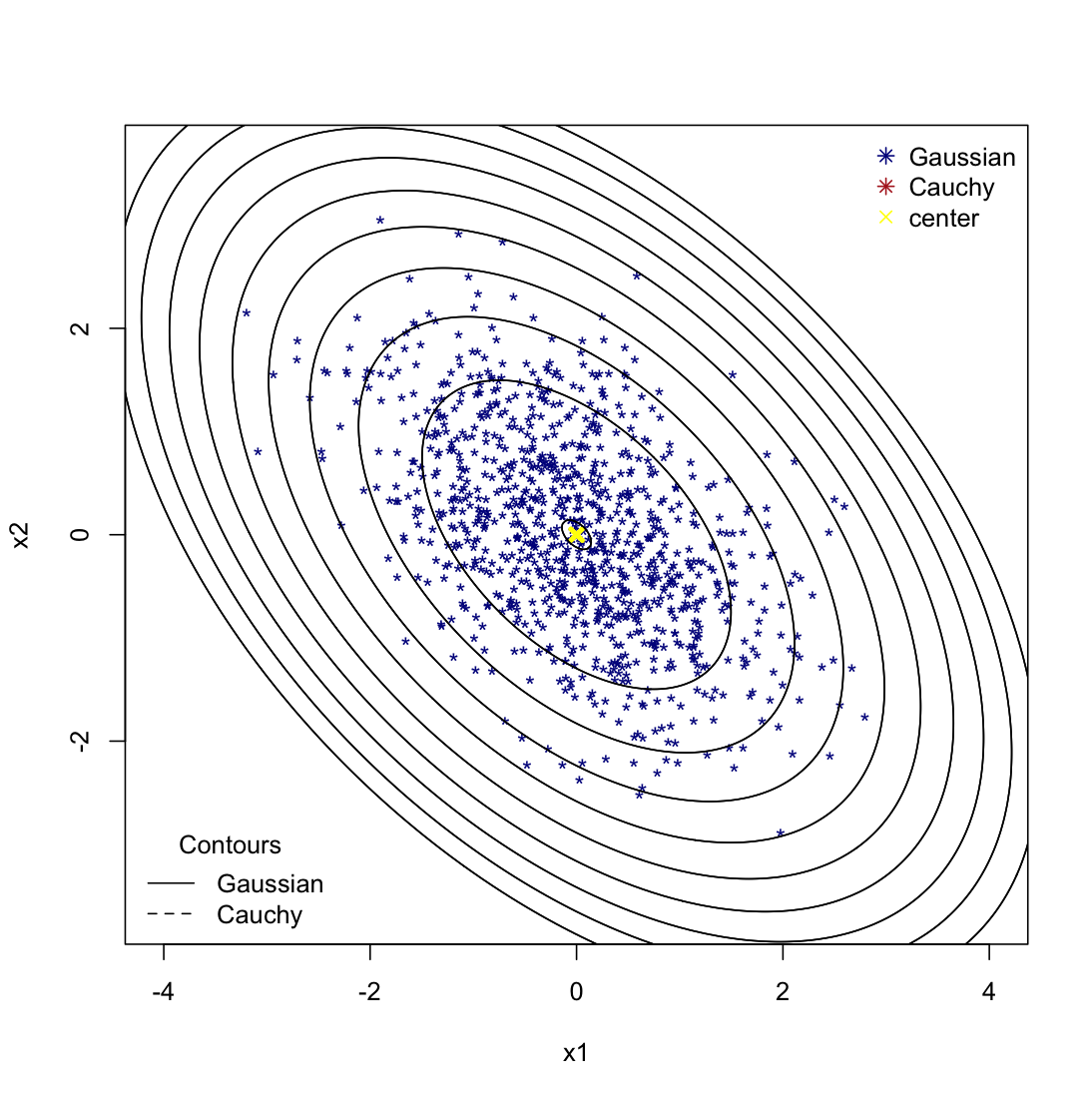}
    \caption{Simulated data}
\end{subfigure}
\hfill
\begin{subfigure}{0.325\textwidth}
    \centering
    \includegraphics[width=\linewidth]{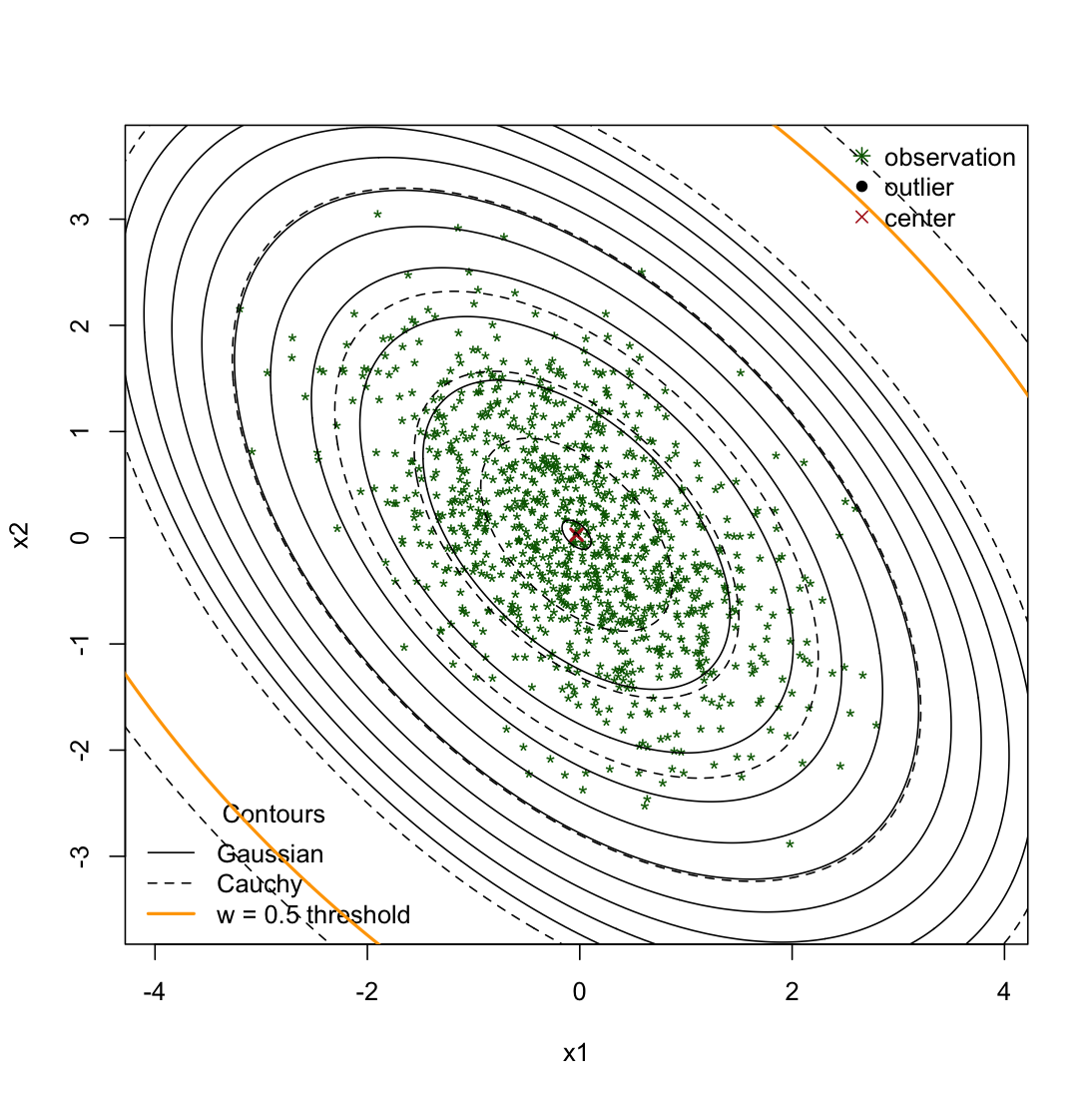}
    \caption{MPVM fit}
\end{subfigure}
\hfill
\begin{subfigure}{0.325\textwidth}
    \centering
    \includegraphics[width=\linewidth]{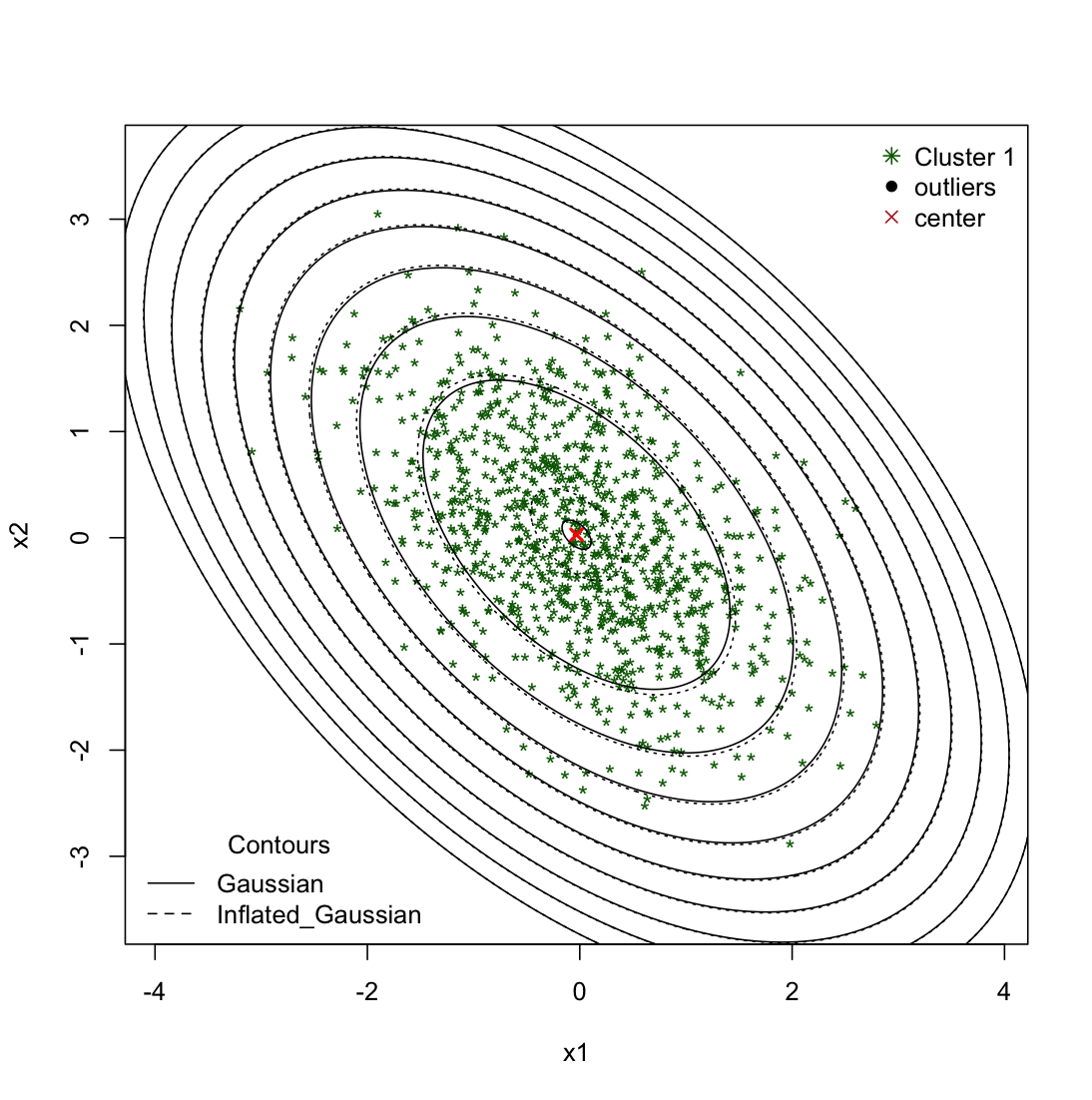}
    \caption{CNM fit}
\end{subfigure}

\caption{Case 1: Visualization of the simulated data and fitted clustering results.}
\label{fig:case1}
\end{figure}

In Case 2, corresponding to a pseudo-Voigt distribution with $\alpha = 0.5$, the MPVM model identifies a single cluster and represents heavy-tailed observations through its Cauchy component. In contrast, the CNM model introduces additional cluster structure (Table~\ref{tab:case2}), resulting in multiple estimated centers.

\begin{table}[htbp]
\centering
\caption{Summary of model estimates and information criteria for case 2.}
\label{tab:case2}
\begin{tabular}{llccc p{4 cm}}
\toprule
Scenario & Model & $\hat{G}$ & BIC & \# Free Par & Estimated $\hat{\mu}$ \\
\midrule
\textbf{Case 2} & MPVM & 1 & 7699.9 & 6&  $(-0.016,0.002)$ \\
& CNM & 2 & 7765.2 & 15&  $(-0.01,0.01), (-0.84,0.76)$ \\
\bottomrule
\end{tabular}
\end{table}

\begin{figure}[htbp]
\centering

\begin{subfigure}{0.325\textwidth}
    \centering
    \includegraphics[width=\linewidth]{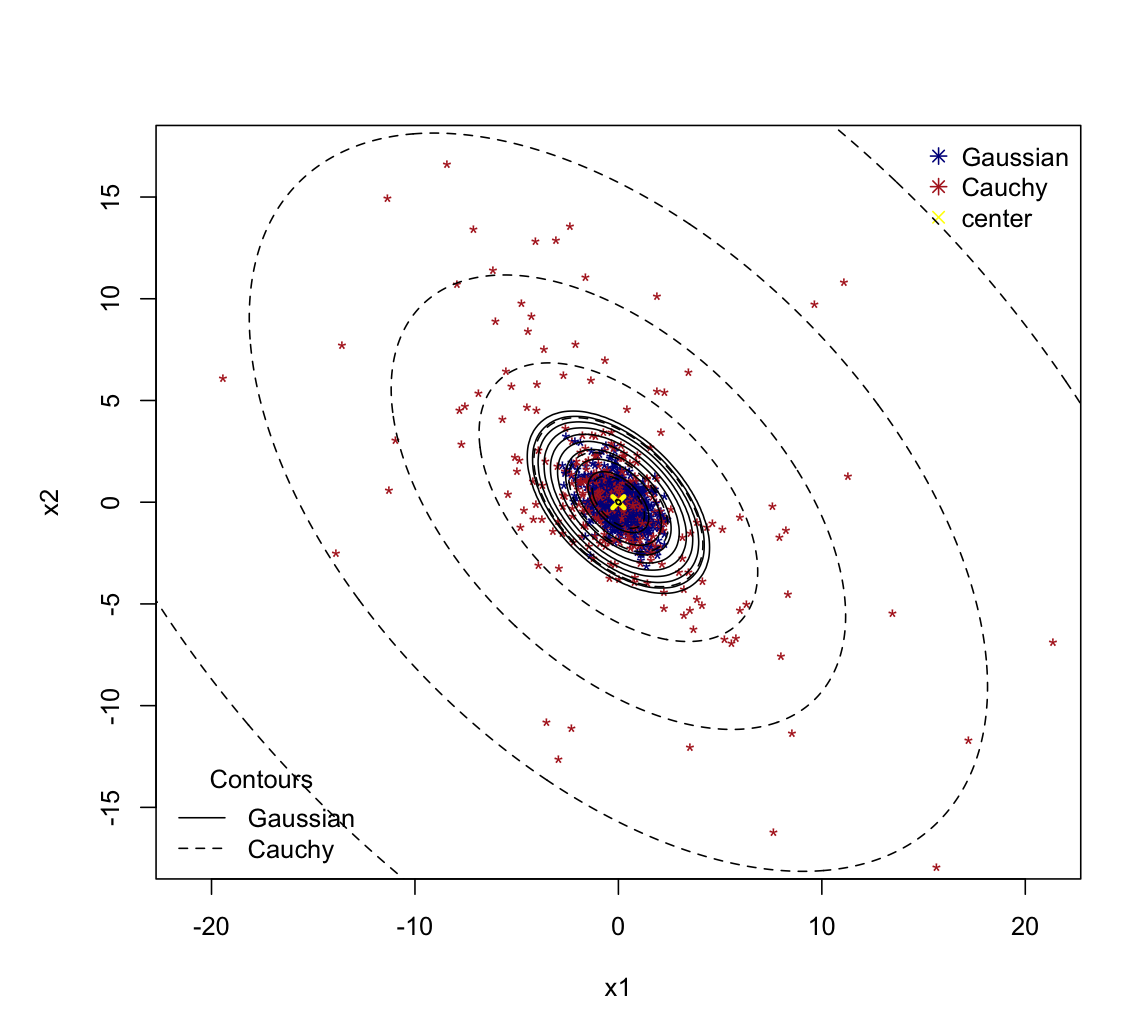}
    \caption{Simulated data}
\end{subfigure}
\hfill
\begin{subfigure}{0.325\textwidth}
    \centering
    \includegraphics[width=\linewidth]{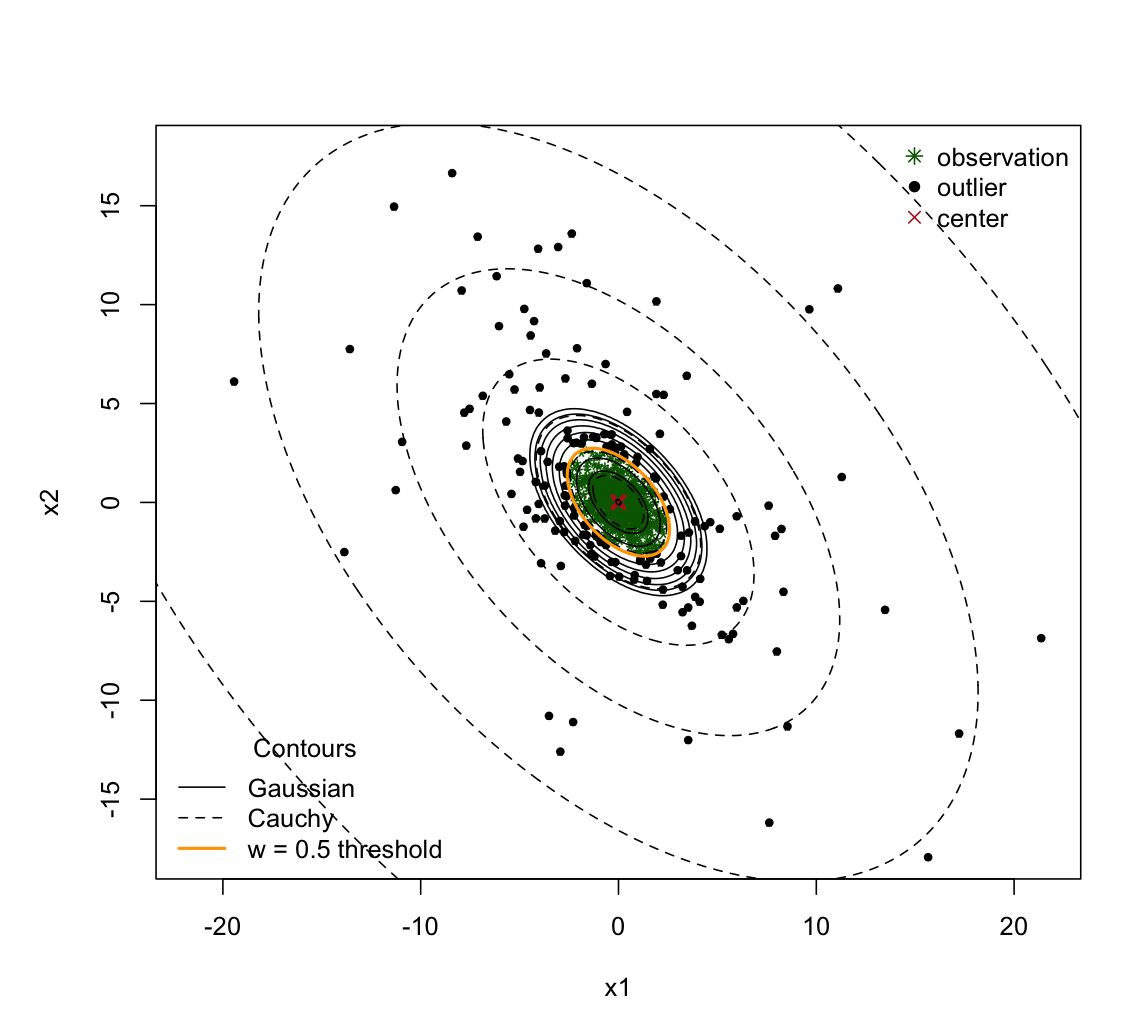}
    \caption{MPVM fit}
\end{subfigure}
\hfill
\begin{subfigure}{0.325\textwidth}
    \centering
    \includegraphics[width=\linewidth]{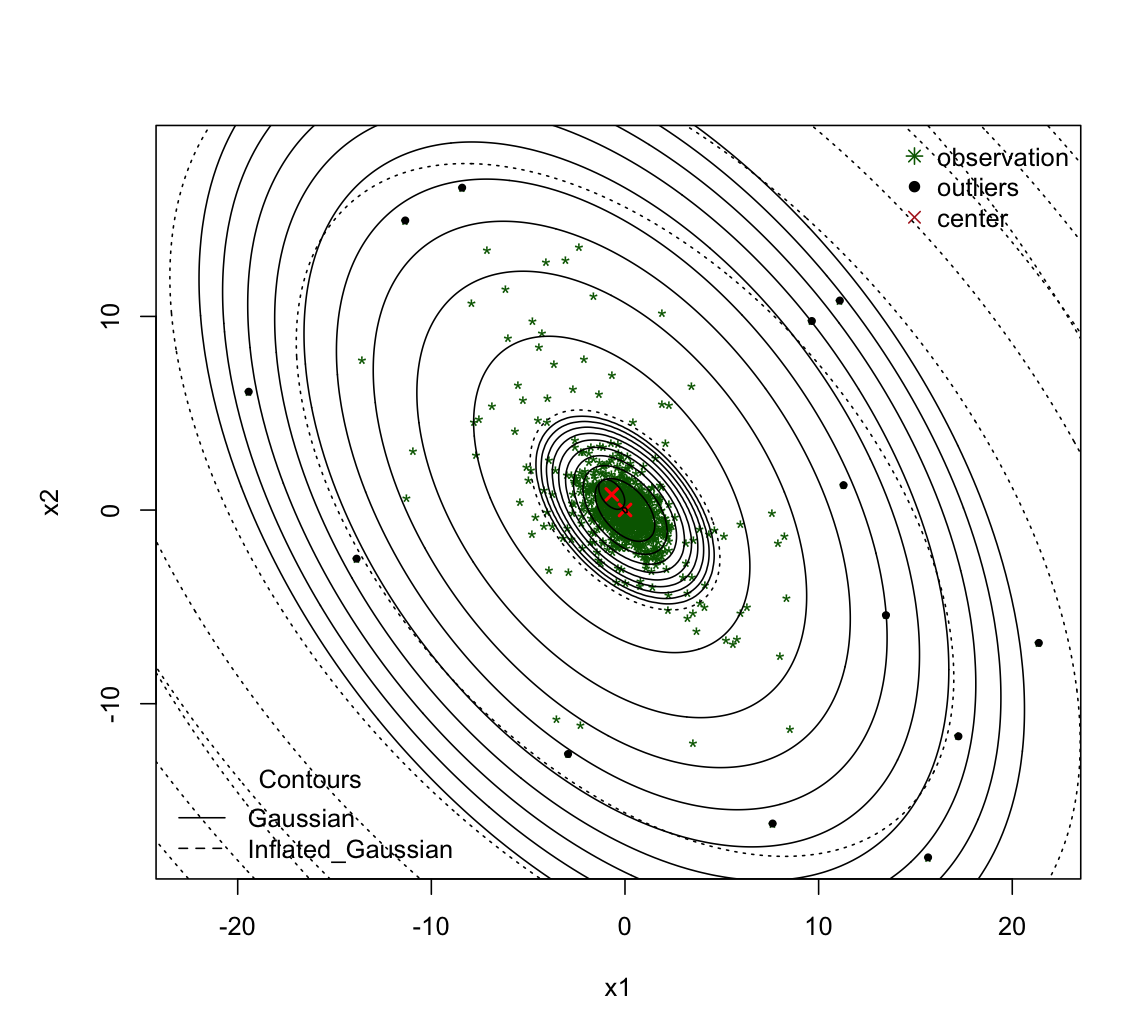}
    \caption{CNM fit}
\end{subfigure}

\caption{Case 2: Visualization of the simulated data and fitted clustering results.}
\label{fig:case2}
\end{figure}

\begin{figure}[htbp]
\centering

\begin{subfigure}{0.325\textwidth}
    \centering
    \includegraphics[width=\linewidth]{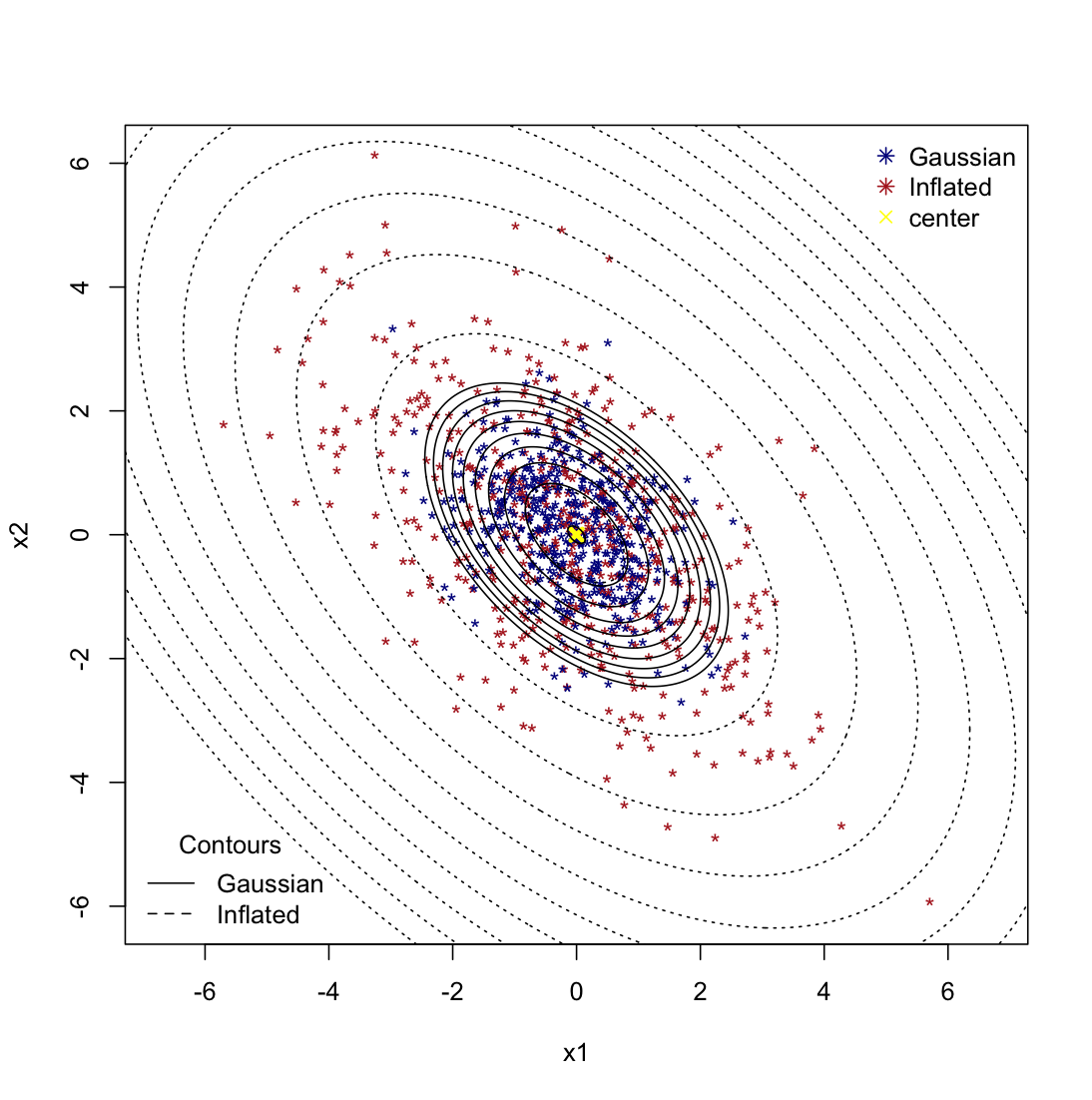}
    \caption{Simulated data}
\end{subfigure}
\hfill
\begin{subfigure}{0.325\textwidth}
    \centering
    \includegraphics[width=\linewidth]{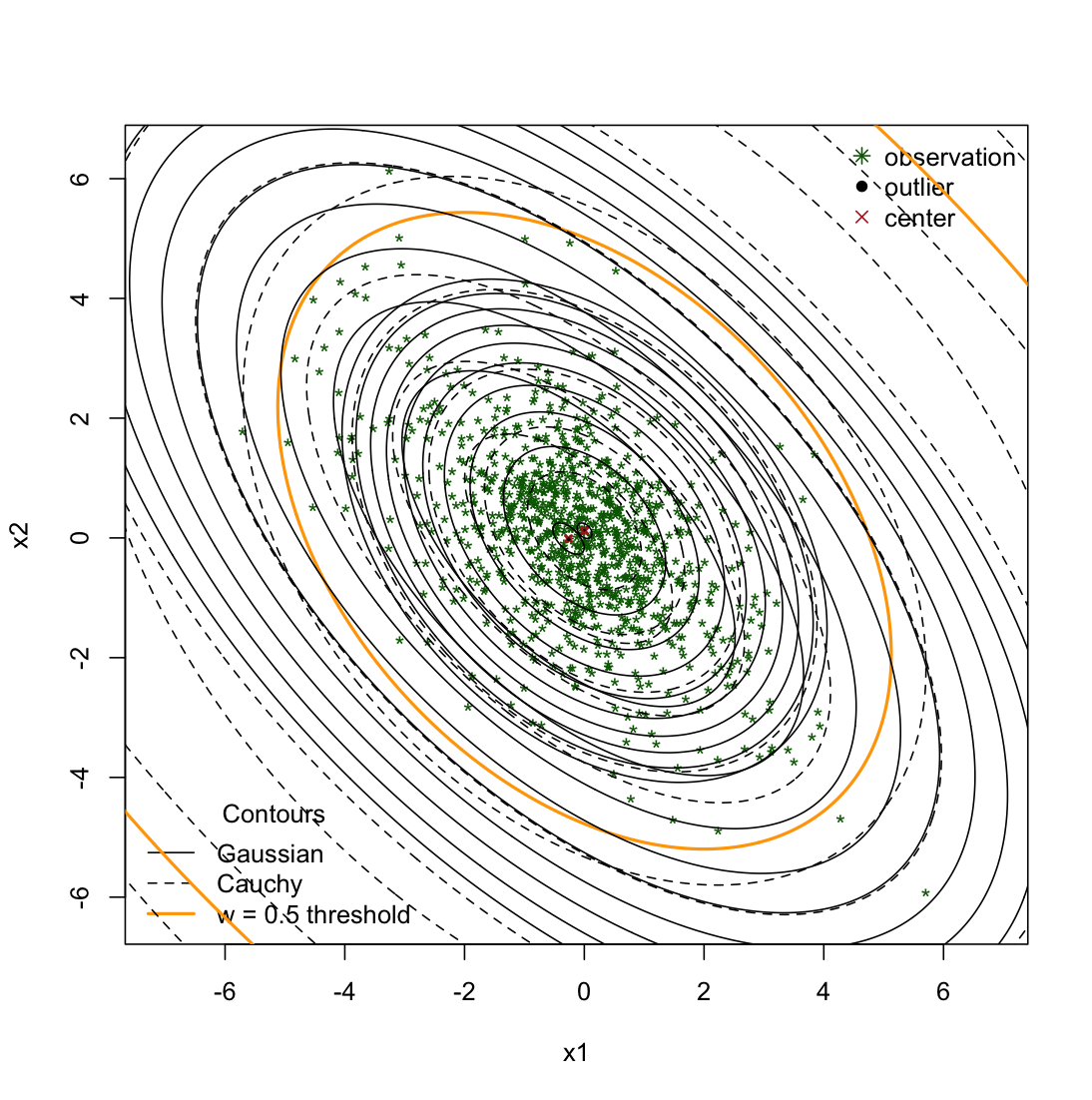}
    \caption{MPVM fit}
\end{subfigure}
\hfill
\begin{subfigure}{0.325\textwidth}
    \centering
    \includegraphics[width=\linewidth]{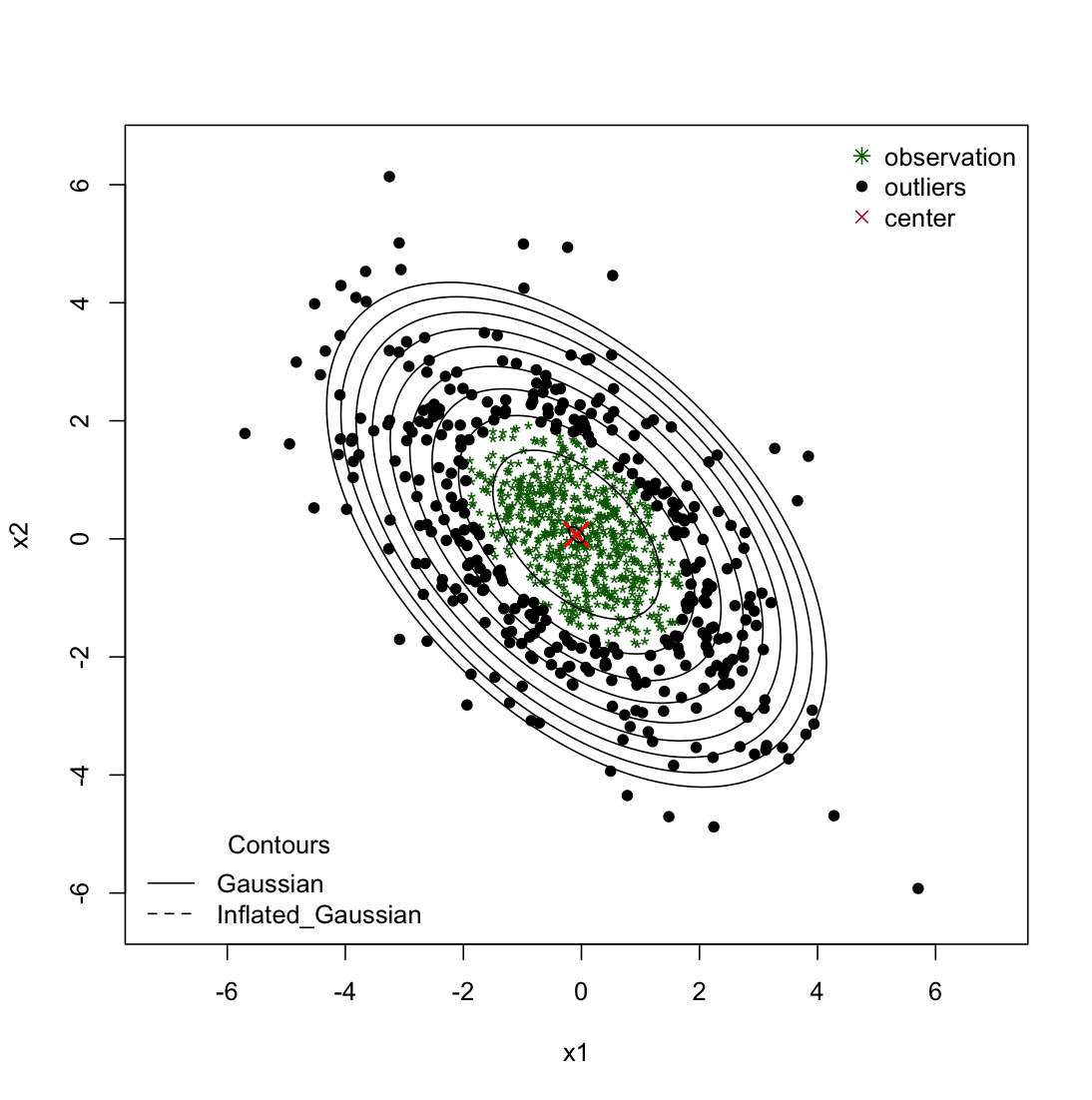}
    \caption{CNM fit}
\end{subfigure}

\caption{Case 3: Visualization of the simulated data and fitted clustering results.}
\label{fig:case3}
\end{figure}

In Case 3, where the data are generated from a contaminated normal distribution with $\alpha = 0.5$, the CNM model identifies a single cluster and captures the inflated variability through outliers, consistent with its formulation (Table~\ref{tab:case3}). Conversely, the MPVM model represents the data using two Gaussian clusters without outliers, indicating that variability is captured through cluster separation rather than within-cluster contamination (Figure~\ref{fig:case3}).

\begin{table}[htbp]
\centering
\caption{Summary of model estimates and information criteria for case 3.}
\label{tab:case3}
\begin{tabular}{llccc p{4 cm}}
\toprule
Scenario & Model & $\hat{G}$ & BIC & \# Free Par & Estimated $\hat{\mu}$ \\
\midrule

\textbf{Case 3} & MPVM & 2 & 6589.3 &  13& $(0.103,0.001), (-0.140,0.135)$ \\
& CNM & 1 & 6551.3 & 7& $(-0.05,0.08)$ \\
\bottomrule
\end{tabular}
\end{table}

\begin{figure}[http]
\centering

\begin{subfigure}{0.325\textwidth}
    \centering
    \includegraphics[width=\linewidth]{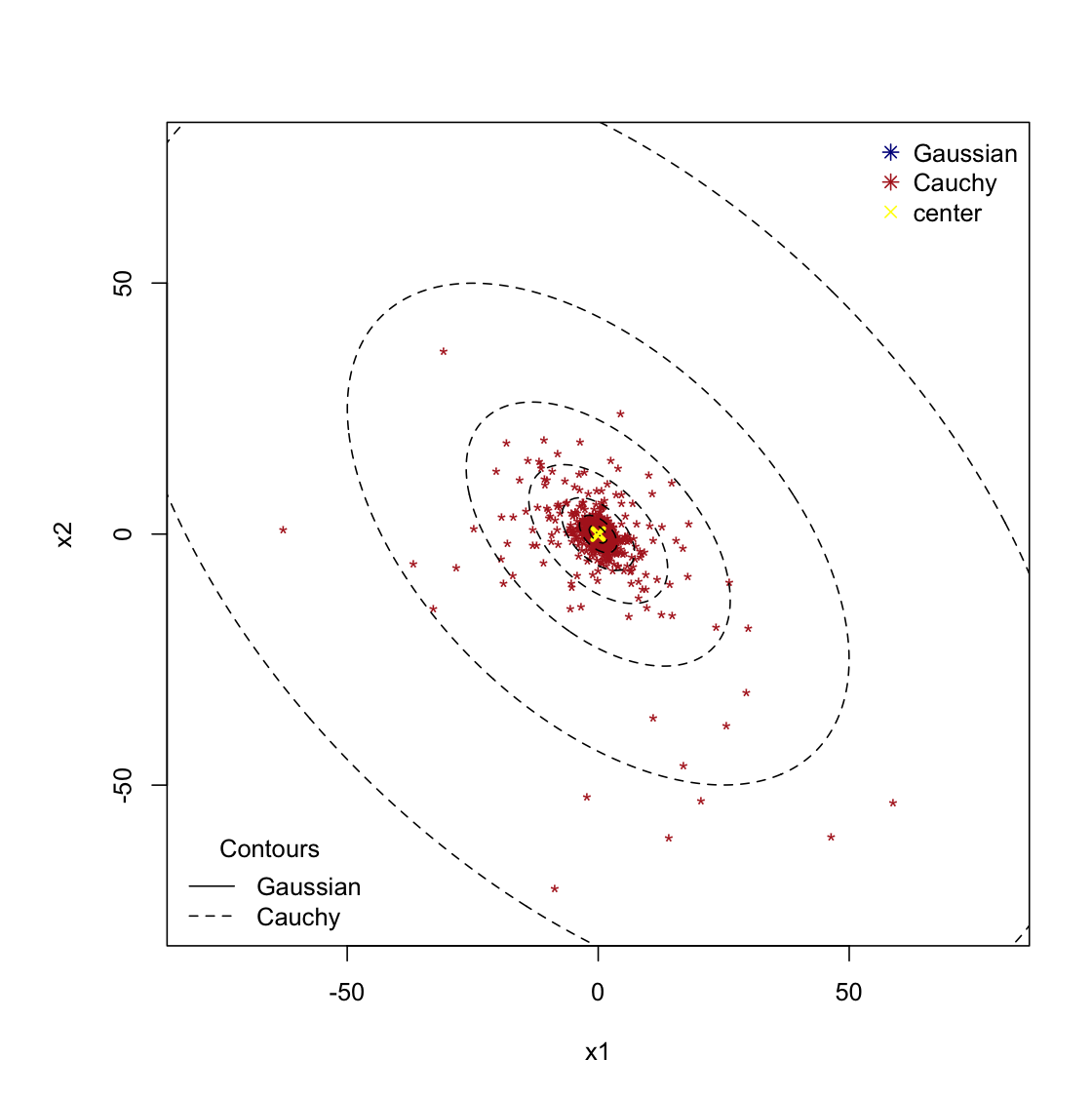}
    \caption{Simulated data}
\end{subfigure}
\hfill
\begin{subfigure}{0.325\textwidth}
    \centering
    \includegraphics[width=\linewidth]{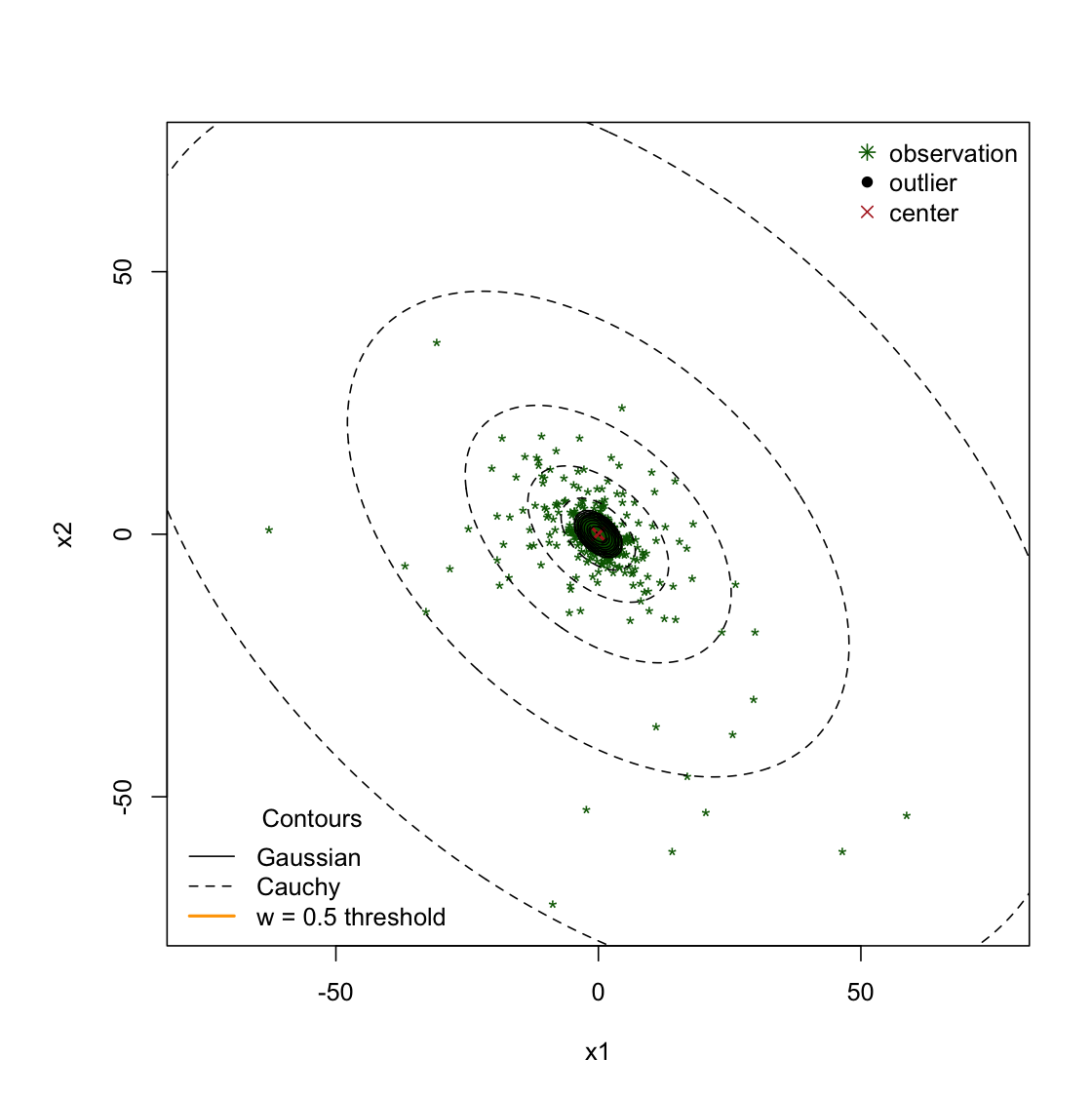}
    \caption{MPVM fit}
\end{subfigure}
\hfill
\begin{subfigure}{0.325\textwidth}
    \centering
    \includegraphics[width=\linewidth]{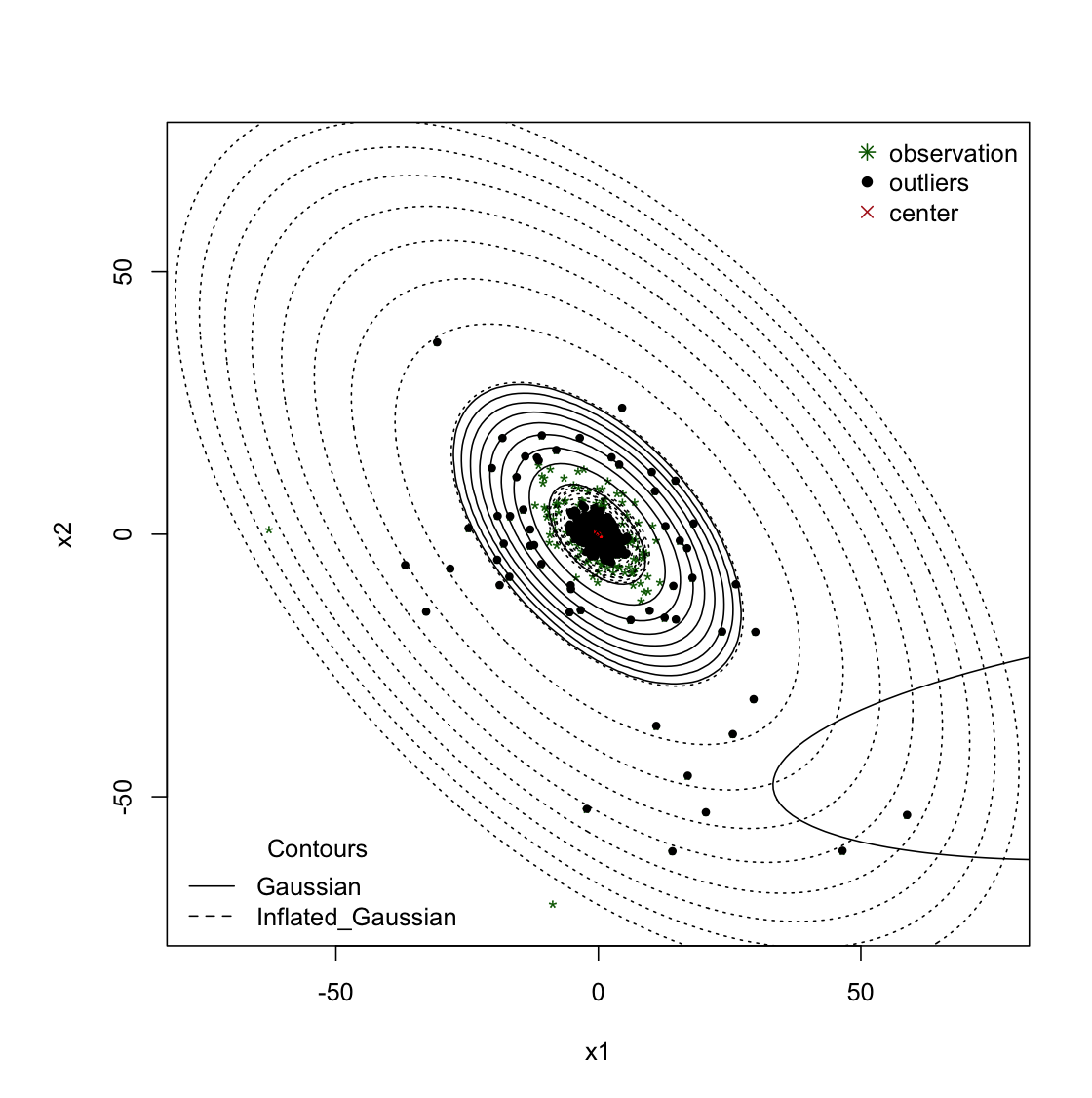}
    \caption{CNM fit}
\end{subfigure}

\caption{Case 4: Visualization of the simulated data and fitted clustering results.}
\label{fig:case4}
\end{figure}

In Case 4, corresponding to a multivariate Cauchy distribution, the MPVM model identifies a single cluster with no outliers, reflecting the fully heavy-tailed nature of the data. The CNM model, however, requires multiple clusters to represent the data structure (Table~\ref{tab:case4} and Figure~\ref{fig:case4}), where extreme observations cause additional cluster formation.

\begin{table}[htbp]
\centering
\caption{Summary of model estimates and information criteria for case 4.}
\label{tab:case4}
\begin{tabular}{llccc p{5.5 cm}}
\toprule
Scenario & Model & $\hat{G}$ & BIC & \# Free Par & Estimated $\hat{\mu}$ \\
\midrule

\textbf{Case 4} & MPVM & 1 & 9482.58 &6 & $(-0.03,0.02)$ \\
& CNM & 3 & 9647.66 & 23 & $(0.01,0.00), (-0.21,-0.02), (125.04,-38.25)$ \\
\bottomrule
\end{tabular}
\end{table}

Overall, these edge-case scenarios demonstrate that the MPVM model provides a stable and unified representation across normal, mixed, and fully heavy-tailed settings. In contrast, the CNM model, due to its reliance on a Gaussian reference structure, may represent heavy-tailed data by introducing additional clusters or through variance inflation within existing clusters in extreme cases.

\subsection{Simulation Study: Cluster Recovery and Outlier Detection}

This section evaluates the performance of the MPVM and CNM models through a comparative simulation study. The primary objective is to investigate how these competing models perform in terms of model fit, assessed using the Bayesian Information Criterion (BIC), and outlier detection accuracy, measured by the F1-score. The models are tested under different distributional settings, specifically Gaussian and $t$-distribution scenarios, to directly assess their capacity to represent cluster structures and identify discordant observations when the true underlying labels are known.

\subsubsection{Data Generation Process}

In both simulation scenarios, data were generated in two dimensions ($p=2$) using two clusters with equal mixing proportions $\boldsymbol{\pi} = (0.5, 0.5)$ and location parameters fixed at $\boldsymbol{\mu}_1 = (0, 3)$ and $\boldsymbol{\mu}_2 = (0, -3)$. Across both regimes, the clusters were assigned distinct covariance structures to provide variation in orientation:
\[
\boldsymbol{\Sigma}_1 =
\begin{pmatrix}
1 & -0.5 \\
-0.5 & 1
\end{pmatrix}, \quad
\boldsymbol{\Sigma}_2 =
\begin{pmatrix}
1 & 0.5 \\
0.5 & 1
\end{pmatrix}.
\]

The data were generated under two distinct distributional settings: a Gaussian scenario ($n=380$), and a $t$-distribution scenario ($n=970$) using $\nu = 8$ degrees of freedom for both clusters.

To provide a ground truth for outlier detection, contamination was introduced by injecting observations sampled from a uniform distribution over the data region (30 points for the Gaussian case and 10 for the $t$-distribution case). Inlier regions were defined using a $99\%$ confidence threshold based on the squared Mahalanobis distance $d^2$. Specifically, an observation was labeled an inlier if
\[
d^2 \leq \chi^2_{p,\,0.99}
\]
for the Gaussian scenario, or
\[
d^2 \leq p \, F_{p,\,\nu;\,0.99}
\]
for the $t$-distribution scenario. Any uniformly generated observation that fell within the defined cluster ellipses was not classified as an outlier.

\subsubsection{Results and Discussion}
The performance of the MPVM and CNM models, averaged over 10 independent repetitions, is summarized in Table~\ref{tab:sim_results}. 

In the Gaussian scenario, both models perform similarly in terms of model fit and outlier detection. The CNM model achieves a lower average BIC, consistent with its Gaussian-based formulation, while the MPVM model attains a marginally higher F1-score, indicating comparable effectiveness in identifying atypical observations.

\begin{figure}[htbp]
    \centering
    \begin{subfigure}{0.45\textwidth}
        \centering
        \includegraphics[width=\linewidth]{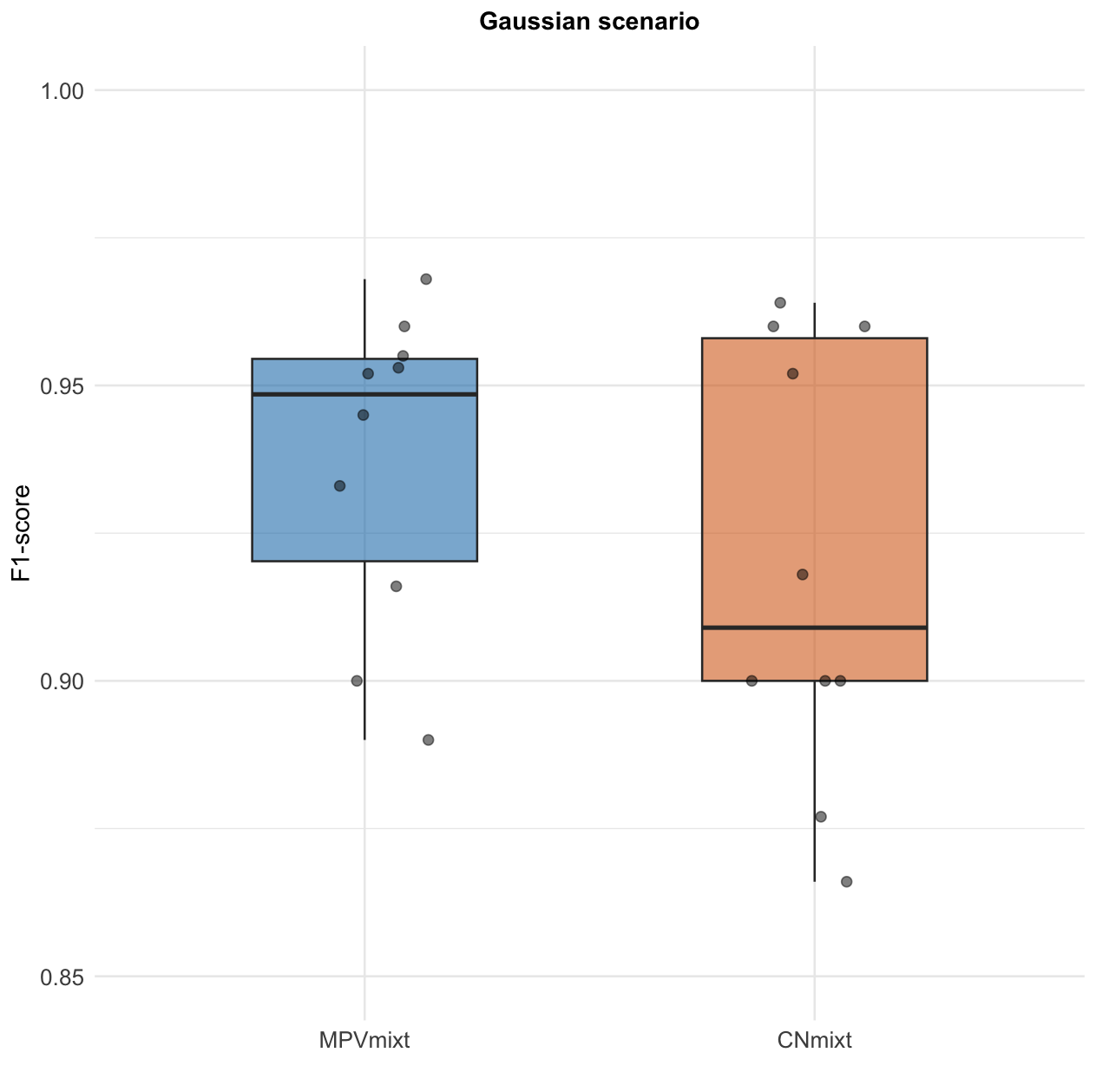}
        \caption{Gaussian Scenario}
    \end{subfigure}
    \hfill
    \begin{subfigure}{0.45\textwidth}
        \centering
        \includegraphics[width=\linewidth]{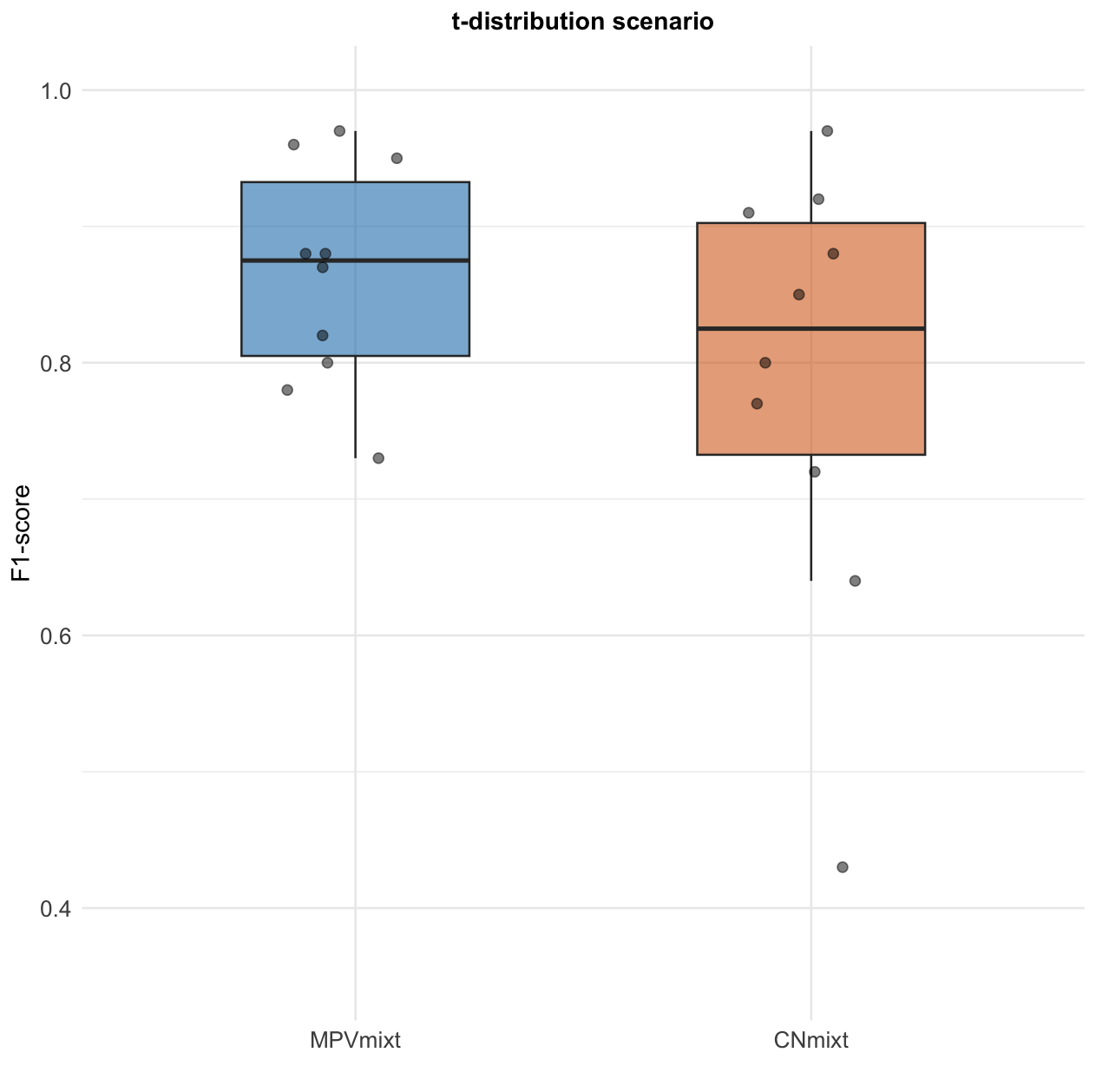}
        \caption{$t$-distribution Scenario}
    \end{subfigure}
    \caption{Distribution of F1-scores across 10 independent repetitions.}
    \label{fig:boxplots}
\end{figure}

In the $t$-distribution scenario, clearer differences emerge. The MPVM model achieves both a lower average BIC and a higher average F1-score, suggesting improved fit and more reliable outlier detection under heavy-tailed conditions. Additionally, the variability of the F1-score across repetitions is lower for MPVM, indicating more stable performance, whereas the CNM model exhibits greater sensitivity to the heavier tails (Figure \ref{fig:boxplots}).

Overall, the results indicate that while both models are suitable for Gaussian data, the MPVM model provides more stable and accurate performance when the underlying distribution deviates from Gaussian assumptions.

\begin{table}[htbp]
\centering
\caption{Average BIC and F1-score over 10 repetitions.}
\label{tab:sim_results}
\small
\setlength{\tabcolsep}{9pt}
\renewcommand{\arraystretch}{1.2}
\begin{tabular}{l  l  cccc}
\toprule
Scenario & Model & Mean BIC & SD & Mean F1 & SD \\
\midrule
\textbf{Gaussian} 
    & MPVM & 3194.81 & 28.8  & \textbf{0.937} & 0.027 \\
    & CNM  & \textbf{3185.12} & 29.4  & 0.920          & 0.037 \\
\midrule
\textbf{$t$-distribution} 
    & MPVM & \textbf{7335.04} & 88.36 & \textbf{0.864} & 0.084 \\
    & CNM  & 7340.72          & 87.60 & 0.819          & 0.136 \\
\bottomrule
\end{tabular}
\end{table}

\subsection{Real Data Analysis}

This section evaluates the performance of the MPVM model using a spectroscopic dataset. The analyses demonstrate the model's capacity to capture biological variability and heterogeneous spectral characteristics that are often absent in controlled simulations. Clustering performance is evaluated against known group labels using the Adjusted Rand Index (ARI), which quantifies the agreement between estimated and true classifications. Model fit is further assessed using the Bayesian Information Criterion (BIC), enabling a comparison of the proposed model with Gaussian mixture model, mixtures of multivariate t-distributions, and contaminated normal mixtures.

\subsubsection{Dataset Description}

The dataset comprises Raman spectroscopy measurements from three human cancer cell lines: H460 (lung cancer) and MCF7 (breast cancer), both analyzed with and without metformin treatment, as well as the radiosensitive LNCaP prostate cancer cell line, which was excluded from the metformin treatment experiment. Each observation corresponds to a Raman spectrum, recording intensity values across a sequence of wavenumbers and reflecting the biochemical composition of the sample. The dataset analyzed in this study was originally collected by \citet{matthews2015radiation} and was used to investigate radiation-induced glycogen accumulation, radioresistance, and the effect of metformin on radiation response in human cancer cell lines.

The original study reported that H460 and MCF7 cells exhibited radiation-induced glycogen accumulation, whereas the radiosensitive LNCaP cell line showed little evidence of this response. Furthermore, metformin reduced radiation-induced glycogen accumulation and increased radiosensitivity in MCF7 cells, whereas it had little effect on H460 cells \citep{matthews2015radiation}.

Building on the findings of \cite{matthews2015radiation}, cluster analyses were conducted using glycogen-associated Raman peaks. The analysis evaluates whether these peaks can differentiate Raman spectra from H460 cells treated with metformin, MCF7 cells treated with metformin, and LNCaP cells.

\subsubsection{Feature Selection}\label{subsubsec:feature selection}
A subset of biologically relevant Raman peaks was selected for the clustering analysis. As reported by \citet{matthews2015radiation}, radiation-induced glycogen accumulation varies among different cancer cell lines. Therefore, the glycogen-associated Raman peaks identified by \citet{Harder2016} were chosen as the primary features for clustering. Specifically, seven glycogen-associated Raman peaks (482, 850, 940, 1042, 1083, 1129, and 1385~cm$^{-1}$) were selected. Additionally, the peak at 1003~cm$^{-1}$, a marker for phenylalanine, was included alongside the glycogen-associated peaks.  

The peak positions reported by \citet{Harder2016} did not always align precisely with those observed in the current dataset. As a result, each Raman peak identified in the literature was visually matched to the nearest corresponding peak in the acquired spectra before feature extraction. The resulting peak positions used in the present study are summarized in Table~\ref{tab:glycogen_peaks}.
\vspace{-.5 cm}
\begin{table}[ht]
\centering
\caption{Reported glycogen-associated Raman peaks and the approximately corresponding peak positions used in the present study.}
\label{tab:glycogen_peaks}
\setlength{\tabcolsep}{9pt}
\begin{tabular}{cccccccc}
\toprule
Reported Peak (cm$^{-1}$) 
& 482 & 850 & 940 & 1042 & 1083 & 1129 & 1385 \\
\midrule
Selected Peak (cm$^{-1}$) 
& 483 & 852 & 936 & 1044 & 1086 & 1125 & 1388 \\
\bottomrule
\end{tabular}
\end{table}

\vspace{-1 cm }
\subsubsection{Clustering of Three Cancer Cell lines}\label{subsubsec: clustering of three cancer cell lines}
To assess the clustering performance of the proposed and comparative models, 990 Raman spectra were selected. This dataset included 330 spectra from H460 cells treated with metformin, 330 spectra from MCF7 cells treated with metformin, and 330 spectra from LNCaP cells. Each Raman spectrum contained measurements at 582 wavenumbers. For clustering, as discussed in \ref{subsubsec:feature selection}, a subset of eight biologically relevant spectral features was chosen. Seven glycogen-associated Raman peaks (482, 850, 940, 1042, 1083, 1129, and 1385~cm$^{-1}$) and one peak at 1003~cm$^{-1}$ as an additional spectral feature.

\begin{figure}[http]
    \centering
    \includegraphics[width=1\linewidth]{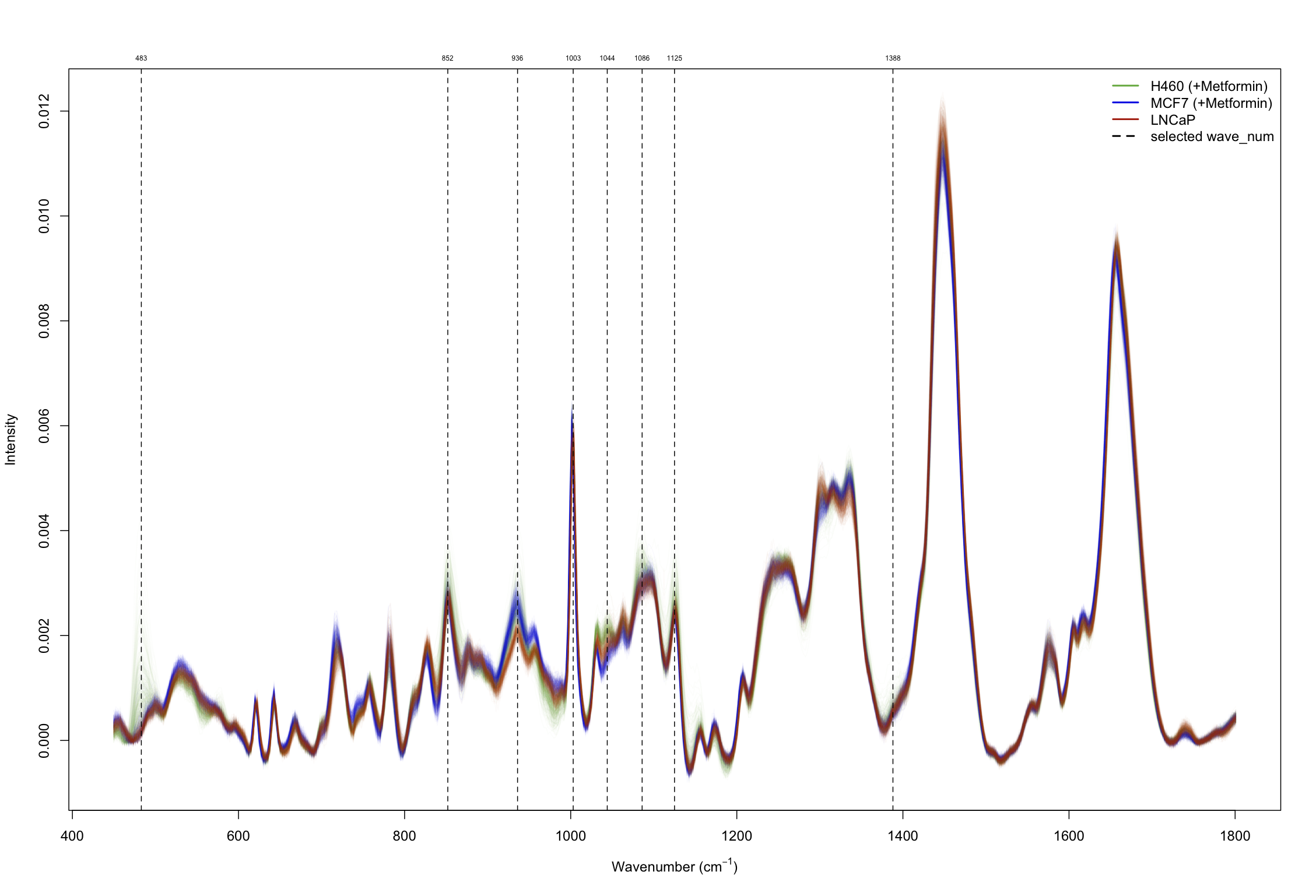}
    \caption{Overlaid Raman spectra for the three cancer cell lines: H460 (+Metformin), MCF7 (+Metformin), and LNCaP. Dashed vertical lines indicate the selected wavenumbers used for feature extraction.}
    \label{fig:raman_spectra}
\end{figure}
\begin{figure}[htbp]
    \centering
    \includegraphics[width=1\linewidth]{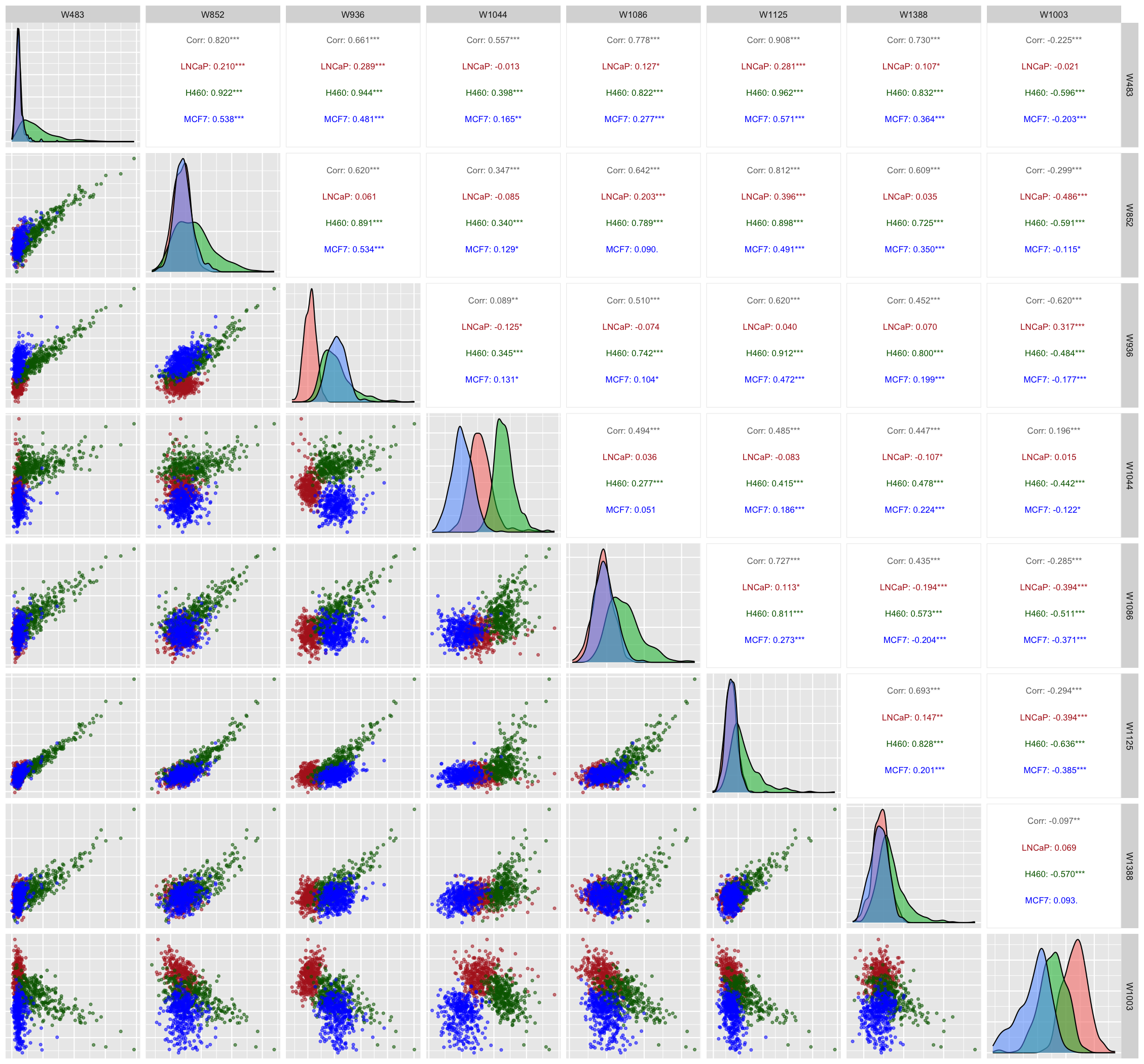}
    \caption{Pairwise scatter plots of the selected wavenumbers for the three cancer cell lines: H460 (+Metformin), MCF7 (+Metformin), and LNCaP (control). Observations are colored by their true labels, illustrating group separation and relationships among the selected spectral features.}
    \label{fig:2_pairwise_cor}
\end{figure}

A reduced dataset was then constructed by retaining the intensity values at the selected wavenumbers, resulting in a 8-dimensional feature representation for each observation. To further examine the structure of the reduced dataset, pairwise scatter plots of the selected features are presented in Figure~\ref{fig:2_pairwise_cor}. These plots indicate moderate separation among the cell lines, with some overlap, reflecting the complexity of the data and motivating the use of flexible mixture models.

The availability of true cell line labels enables external validation of clustering performance. In addition to model-based criteria such as BIC, this allows direct assessment of how effectively the competing models recover underlying group membership.

\subsubsection{Model Specification}

Four model-based clustering approaches were applied to the reduced dataset: the proposed multivariate pseudo-Voigt mixture model (MPVM), the contaminated normal mixture model (CNM), a mixture of $t$-distributions, and a Gaussian mixture model (GMM).

For MPVM, CNM, and GMM, the most general covariance structure (VVV) was used, allowing each component to have an unconstrained covariance matrix. For the mixture of $t$-distributions, the fully unconstrained structure (UUUU) was adopted to ensure comparable flexibility across models.

Each model was fitted with the number of components $G$ ranging from 1 to 5. Initialization was performed using the modified trimmed $k$-means strategy described in Section~\ref{subsec:initialization}, which was applied consistently across all models to ensure fair comparison.

Model selection relied on the Bayesian Information Criterion (BIC), with the optimal model identified as the one that minimized BIC. Since certain software implementations report BIC using the alternative convention \(2\ell(\hat{\boldsymbol{\Psi}}) - q\log n\), these values were converted to the \(-2\ell(\hat{\boldsymbol{\Psi}}) + q\log n\) formulation adopted in this study to enable consistent model comparisons.

\subsubsection{Results}

The best Bayesian Information Criterion (BIC) values for each model are presented in Table~\ref{tab:bic_summary} to facilitate comparison of clustering results. Three of four models identify three clusters, indicating a consistent structural pattern, whereas the GMM selected four clusters. Among the evaluated approaches, the MPVM model achieves the lowest BIC value of $-127736.4$, indicating the most favorable overall fit.
\begin{table}[htbp]
\centering
\caption{Selected number of clusters, corresponding BIC values, and BIC differences relative to MPVM. Lower BIC values indicate better model fit.}
\label{tab:bic_summary}
\setlength{\tabcolsep}{10pt}
\begin{tabular}{lcccc}
\toprule
 & CNM & GMM & $t$-mixture & MPVM\\
\midrule
Selected number of clusters & 3 & 4 & 3 & 3\\
BIC & $-127715.9$ & $-127651.8$ & $-127730.1$ & $-127736.4$\\
$\Delta\mathrm{BIC}$ relative to MPVM & 20.5 & 84.6 & 6.3 & 0\\
\bottomrule
\end{tabular}
\end{table}

The BIC differences can be interpreted as approximations to twice the log Bayes factor \citep{KassRaftery1995}. Values between 6 and 10 indicate strong evidence, while values greater than 10 indicate very strong evidence against the model with the larger BIC. Compared to MPVM, the $\Delta\mathrm{BIC}$ values for the CNM, GMM, and $t$-mixture are 20.5, 84.6, and 6.3, respectively. These findings provide very strong evidence in favour of MPVM over the CNM and GMM, and strong evidence in favour of MPVM over the $t$-mixture.

The relationship between the identified clusters and true cell line labels is assessed by cross-tabulating the true groups with the estimated cluster assignments, as shown in Table~\ref{tab:classification}.


\begin{table}[htbp]
\centering
\footnotesize
\caption{Cross-tabulation of true labels versus estimated cluster assignments.}
\label{tab:classification}

\setlength{\tabcolsep}{6.5pt} 

\begin{tabular}{c|ccc|cccc|ccc|ccc}
\toprule
& \multicolumn{3}{c|}{CNM} 
& \multicolumn{4}{c|}{GMM} 
& \multicolumn{3}{c|}{$t$-mixture} 
& \multicolumn{3}{c}{MPVM} \\
\cmidrule(lr){2-4} \cmidrule(lr){5-8} \cmidrule(lr){9-11} \cmidrule(lr){12-14}
True & 1 & 2 & 3 & 1 & 2 & 3 & 4 & 1 & 2 & 3 & 1 & 2 & 3 \\
\midrule
LNCaP & 357 & 0 & 3 & 347 & 0 & 11 & 2 & 357 & 0 & 3 & 358 & 0 & 2 \\
H460  & 5 & 0 & 355 & 3 & 0 & 196 & 161 & 4 & 0 & 356 & 4 & 0 & 356 \\
MCF7  & 1 & 357 & 2 & 1 & 356 & 3 & 0 & 1 & 357 & 2 & 1 & 357 & 2 \\
\midrule
\end{tabular}
\end{table}


The results indicate that the MPVM, CNM, and multivariate $t$-mixture models effectively recover the three cell lines with minimal misclassification, assigning nearly all observations from each cell line to a single cluster. In contrast, the GMM divides the H460 cell line into two distinct clusters, leading to the identification of an additional cluster beyond the established biological groups. The three models that successfully recover the underlying three-cluster structure each incorporate mechanisms to accommodate heavy-tailed observations, either through a heavy-tailed component ($t$-mixture and MPVM) or an explicit contamination model (CNM). These findings suggest that addressing deviations from Gaussianity enhances the accurate characterization of the underlying structure in Raman spectroscopy data.

Clustering performance was further evaluated using the adjusted Rand index (ARI), with the results summarized in Table~\ref{tab:ari}.


\begin{table}[htbp]
\centering
\small
\setlength{\tabcolsep}{13pt}
\caption{Adjusted Rand Index (ARI) values for competing models.}
\label{tab:ari}
\setlength{\tabcolsep}{16 pt}
\begin{tabular}{lcccc}
\toprule
Model  & CNM & GMM & $t$-mixture & MPVM\\
\midrule
ARI   & 0.969   & 0.825  & 0.972 & 0.975 \\
\bottomrule
\end{tabular}
\end{table}


The ARI values demonstrate strong agreement between the estimated cluster assignments and the true cell line labels for all robust mixture models. The proposed MPVM model achieves the highest ARI of 0.975, followed by the multivariate $t$-mixture (0.972) and the CNM (0.969). In comparison, the GMM gives a lower ARI of 0.825, which is attributable to its tendency to divide the H460 cell line into two distinct clusters. Collectively, these findings indicate that the three robust mixture models accurately recover the underlying biological groups, with the proposed MPVM exhibiting the highest agreement with the known class labels.

In addition to clustering performance, the MPVM and CNM models explicitly identify potential outliers through their respective detection mechanisms. As shown in Table \ref{tab:outliers_detailed}, MPVM identifies 11 number of observations as outliers across all groups. In contrast, CNM classifies 78 observations as atypical.


\begin{table}[htbp]
\centering
\small
\caption{Cross-tabulation of true groups versus estimated clusters, including the number of observations identified as outlier by MPVM and CNM.}
\label{tab:outliers_detailed}
\begin{tabular*}{\textwidth}{@{\extracolsep{\fill}}c|cccc|cccc}
\toprule
 & \multicolumn{4}{c|}{MPVM} & \multicolumn{4}{c}{CNM} \\
\cmidrule(lr){2-5} \cmidrule(lr){6-9}
True Group & 1 & 2 & 3 & Outlier & 1 & 2 & 3 & Outlier \\
\midrule
1 (LNCaP)  & 355 & 0 & 2 & 3 & 327 & 0 & 3 & 30  \\
2 (H460)   & 4 & 0 & 353 & 3 & 4 & 0 & 333 & 23 \\
3 (MCF7)   & 1 & 354 & 0 & 5 & 1 & 334 & 0 & 25 \\
\bottomrule
\end{tabular*}
\end{table}


To further analyze the results in \ref{tab:outliers_detailed}, the estimated parameters from both models were compared. Both models produced nearly nearly identical estimates of cluster centers $(\boldsymbol{\mu})$ and mixing proportions $(\boldsymbol{\pi})$; however, they differed substantially in their estimated components weights. The estimated $\boldsymbol{\alpha}$ parameters explains the difference in outlier detection reported in \ref{tab:outliers_detailed}. The CNM model estimated $\boldsymbol{\alpha} = (0.84, 0.85, 0.85)$, indicating that approximately $15\%$ of the observations within each cluster were assigned to the contaminated component. In contrast, the MPVM model estimated $\boldsymbol{\alpha} = (0.97, 0.96, 0.95)$, consistent with the substantially lower number observations identified as atypical. These parameter estimates (reported in Appendix \ref{secB}) suggest that the MPVM accommodates tail observations within the primary cluster structure, whereas the CNM model is more likely to classify such observations as outliers.

\section{Discussion}\label{sec: discussion}

This paper introduces the MPVM model, a mixture modeling framework based on the multivariate pseudo-Voigt distribution. The approach integrates a Gaussian component with a heavy-tailed Cauchy component within each cluster, enabling flexible modeling of both light-tailed and heavy-tailed behaviors. In contrast to contamination-based methods, outlier identification in MPVM model is not determined solely by membership in a specific contamination component. Instead, the model combines a generative representation of heavy-tailed behavior with a geometric dominance criterion derived from the relative Cauchy to Gaussian ratio. Consequently, observations are classified according to their position within the central or tail regions of a cluster, which allows heavy-tailed observations to be incorporated into the cluster structure while maintaining a principled mechanism for identifying atypical points.

The behavioral study further demonstrates the impact of the MPVM model's structural formulation on clustering behavior. The MPVM model is capable of representing heavy-tailed distributions within a single cluster, whereas competing approaches often introduce additional clusters to account for extreme observations. Nevertheless, the findings also reveal certain limitations. In particular, when contamination exhibits a variance-inflated Gaussian structure, the MPVM model may interpret the increased variability as evidence of cluster separation instead of within-cluster dispersion.

The simulation studies highlight several key characteristics of the MPVM model. In Gaussian settings, MPVM performs comparably to the contaminated normal mixture, demonstrating that its additional flexibility does not compromise performance when heavy tails are absent. In contrast, under heavy-tailed scenarios, the MPVM model exhibits greater stability and robustness. Notably, it reduces the tendency to over-segment the data, a behavior commonly observed in Gaussian-based frameworks when heavy-tailed observations are present.

Analysis of the real data indicates that all four models capture the underlying structure of the Raman spectroscopy data. The MPVM, contaminated normal mixture, and multivariate t-mixture models each recover the three biological groups with near-perfect agreement, whereas the Gaussian mixture model over-partitions the H460 cell line into an additional cluster. Among the evaluated approaches, the MPVM model attains both the lowest Bayesian Information Criterion (BIC) and the highest adjusted Rand index, demonstrating superior overall clustering performance. These results suggest that, in this application, allowing for departures from Gaussian assumptions provides a more flexible representation of the observed cluster structure. 

Future research may extend the model to include parsimonious covariance structures and asymmetric component distributions.





\begin{appendices}

\section{Existence and Uniqueness of the Dominance Threshold}\label{secA1}

This section establishes the existence and uniqueness of a relative dominance threshold for the shared-scale multivariate pseudo-Voigt (MPV) model. The normalized log-ratio between the Cauchy and Gaussian components is shown to depend solely on the squared Mahalanobis distance from the common location parameter. Analysis of this function demonstrates that a unique threshold exists beyond which the Cauchy component dominates the Gaussian component. As a result, the boundary separating the central and tail regions is defined by a single ellipsoidal contour, and observations beyond this boundary remain in the tail region as their Mahalanobis distance increases.

Suppose that the Gaussian and Cauchy densities have a common location parameter $\boldsymbol{\mu}$ and a common scale matrix $\boldsymbol{\Sigma}$. Let
\[
\delta(\mathbf{x}) = (\mathbf{x}-\boldsymbol{\mu})^\top \boldsymbol{\Sigma}^{-1} (\mathbf{x}-\boldsymbol{\mu})
\]
represent the squared Mahalanobis distance from $\boldsymbol{\mu}$. We define the relative dominance ratio as follows:
\[
R(\mathbf{x}) = \frac{(1-\alpha)C(\mathbf{x}\mid\boldsymbol{\mu},\boldsymbol{\Sigma})}{\alpha G(\mathbf{x}\mid\boldsymbol{\mu},\boldsymbol{\Sigma})}, \qquad \alpha\in(0,1),
\]
and define the normalized log-ratio as
\[
r_g(\mathbf{x}) = \log\left( \frac{R(\mathbf{x})}{R(\boldsymbol{\mu})} \right).
\]

Given the shared-scale Gaussian and Cauchy densities, we have
\[
G(\mathbf{x}\mid\boldsymbol{\mu},\boldsymbol{\Sigma}) = (2\pi)^{-p/2} |\boldsymbol{\Sigma}|^{-1/2} \exp\left\{-\frac{1}{2}\delta(\mathbf{x})\right\},
\]
and for the Cauchy density,
\[
C(\mathbf{x}\mid\boldsymbol{\mu},\boldsymbol{\Sigma}) = k_p |\boldsymbol{\Sigma}|^{-1/2} \left(1+\delta(\mathbf{x})\right)^{-(p+1)/2},
\]
where $k_p$ is the normalizing constant for the multivariate Cauchy density. Therefore, the ratio can be written as:
\[
\frac{R(\mathbf{x})}{R(\boldsymbol{\mu})} = \frac{C(\mathbf{x}\mid\boldsymbol{\mu},\boldsymbol{\Sigma})}{C(\boldsymbol{\mu}\mid\boldsymbol{\mu},\boldsymbol{\Sigma})} \frac{G(\boldsymbol{\mu}\mid\boldsymbol{\mu},\boldsymbol{\Sigma})}{G(\mathbf{x}\mid\boldsymbol{\mu},\boldsymbol{\Sigma})}.
\]

Since $\delta(\boldsymbol{\mu})=0$ at the center, it follows that
\[
\frac{R(\mathbf{x})}{R(\boldsymbol{\mu})} = \left(1+\delta(\mathbf{x})\right)^{-(p+1)/2} \exp\left\{\frac{1}{2}\delta(\mathbf{x})\right\}.
\]
Taking the natural logarithm yields
\[r_g(\mathbf{x}) = \frac{1}{2}\delta(\mathbf{x}) - \frac{p+1}{2} \log\left(1+\delta(\mathbf{x})\right).
\]
Thus, $r_g(\mathbf{x})$ depends on $\mathbf{x}$ solely through the scalar metric $\delta(\mathbf{x})$. 

Now, define the univariate function
\[
m(\delta) = \frac{1}{2}\delta - \frac{p+1}{2}\log(1+\delta), \qquad \delta\ge 0.
\]
It immediately follows that $m(0)=0$. Differentiating $m(\delta)$ with respect to $\delta$ gives
\[
m'(\delta) = \frac{1}{2} - \frac{p+1}{2(1+\delta)} = \frac{\delta-p}{2(1+\delta)}.
\]
Therefore, $m'(\delta) < 0$ on $(0,p)$ and $m'(\delta) > 0$ on $(p,\infty)$, meaning that $m$ decreases on $(0,p)$ and increases on $(p,\infty)$. Consequently, $m$ attains a unique global minimum at $\delta=p$.

Since $m(0)=0$ and $m$ is strictly decreasing on $(0,p)$, we have
\[
m(\delta)<0 \qquad \text{for all } 0<\delta\le p.
\]
Furthermore, because the linear term dominates the logarithmic term as distance grows, we find
\[
\lim_{\delta\to\infty}m(\delta)=+\infty.
\]
Since $m(p) < 0$, $\lim_{\delta\to\infty}m(\delta)=+\infty$, and $m$ is strictly increasing on the interval $(p,\infty)$, the Intermediate Value Theorem guarantees that the equation
\[
m(\delta)=0
\]
has exactly one unique nonzero solution, which we denote by $\delta^\star>p$.

As a result, we establish that
\[
r_g(\mathbf{x})>0 \quad \Longleftrightarrow \quad R(\mathbf{x})>R(\boldsymbol{\mu})
\]
for all observations satisfying
\[
\delta(\mathbf{x})>\delta^\star.
\]

Therefore, in the shared-scale case, the relative dominance threshold corresponds precisely to the unique ellipsoidal boundary
\[
(\text{\bf x}-\boldsymbol{\mu})^\top \boldsymbol{\Sigma}^{-1} (\mathbf{x}-\boldsymbol{\mu}) = \delta^\star.
\]
Any observation whose distance exceeds this boundary is guaranteed to remain in the tail region for all larger Mahalanobis distances.

\section{Detailed Parameter Estimates for Real Data Analysis}\label{secB}

\begin{table}[ht]
\centering
\caption{Estimated Mixing Proportions ($\pi$) and Component Weights ($\alpha$).}
\label{tab:alpha_pi_estimates}
\begin{tabular}{llccc}
\toprule
Parameter & Model & Cluster 1 & Cluster 2 & Cluster 3 \\
\midrule
$\pi$ (Prior) & MPVM & 0.3350 & 0.3311 & 0.3339 \\
              & CNM  & 0.3352 & 0.3310 & 0.3338 \\
              & \textit{True Value} & \textit{0.3333} & \textit{0.3333} & \textit{0.3333} \\
\midrule
$\alpha$ (Weight) & MPVM & 0.9786 & 0.9648 & 0.9463 \\
                  & CNM  & 0.8409 & 0.8553 & 0.8556 \\
\bottomrule
\end{tabular}
\end{table}

\begin{table}[ht]
\centering
\caption{Estimated mean vectors ($\boldsymbol{\mu}$) for the MPVM and CNM models.}
\label{tab:mu_estimates}
\begin{tabular}{l|cc|cc|cc}
\toprule
& \multicolumn{2}{c|}{Cluster 1} &
\multicolumn{2}{c|}{Cluster 2} &
\multicolumn{2}{c}{Cluster 3} \\
\cmidrule(lr){2-3}
\cmidrule(lr){4-5}
\cmidrule(lr){6-7}
Feature & MPVM & CNM & MPVM & CNM & MPVM & CNM \\
\midrule
X.1 & 0.000212 & 0.000211 & 0.000229 & 0.000229 & 0.000805 & 0.000805 \\
X.2 & 0.002673 & 0.002675 & 0.002704 & 0.002703 & 0.002876 & 0.002876 \\
X.3 & 0.002046 & 0.002045 & 0.002544 & 0.002545 & 0.002530 & 0.002530 \\
X.4 & 0.001833 & 0.001832 & 0.001606 & 0.001606 & 0.002125 & 0.002124 \\
X.5 & 0.002963 & 0.002965 & 0.003012 & 0.003011 & 0.003198 & 0.003198 \\
X.6 & 0.002355 & 0.002354 & 0.002363 & 0.002363 & 0.002613 & 0.002613 \\
X.7 & 0.000571 & 0.000571 & 0.000552 & 0.000553 & 0.000665 & 0.000665 \\
X.8 & 0.005669 & 0.005665 & 0.004948 & 0.004951 & 0.005276 & 0.005272 \\
\bottomrule
\end{tabular}
\end{table}




\end{appendices}

\newpage
\bibliographystyle{sn-chicago}

\bibliography{sn-bibliography}

\end{document}